\documentclass[preprint,preprintnumbers,nofootinbib]{revtex4-1}

\usepackage[normalem]{ulem}
\usepackage{graphicx}
\usepackage{dcolumn}
\usepackage{bm}
\usepackage{amsmath}
\usepackage{amsfonts} 
\usepackage{latexsym}
\usepackage{bbm}
\usepackage{color}
\usepackage{amssymb}
\usepackage{amsthm}
\usepackage{epsfig}
\usepackage{physics}
\usepackage{hyperref}
\usepackage{comment}
\usepackage{ulem}
\usepackage{tikz-feynhand}
\hypersetup{
setpagesize=false,
 bookmarksnumbered=true,
 bookmarksopen=true,
 colorlinks=true,
 linkcolor=blue,
 citecolor=red,
}

\begin{document}

\preprint{\textbf{OCHA-PP-380}}

\title{Degenerate scalar scenario of two Higgs doublet model with a complex singlet scalar}

\author{Gi-Chol Cho$^1$}
\email{cho.gichol@ocha.ac.jp}

\author{Chikako Idegawa$^{1,2}$}
\email{c.idegawa@hep.phys.ocha.ac.jp, idegawa@mail.sysu.edu.cn}

\affiliation{$^1$ Department of Physics, Ochanomizu University, Tokyo 112-8610, Japan}

\affiliation{$^2$ MOE Key Laboratory of TianQin Mission,
TianQin Research Center for Gravitational Physics \& School of Physics and Astronomy,
Frontiers Science Center for TianQin,
Gravitational Wave Research Center of CNSA,
Sun Yat-sen University (Zhuhai Campus), Zhuhai 519082, China}

\bigskip

\date{\today}
\begin{abstract}
We study the two Higgs doublet model with a complex singlet scalar whose imaginary part acts as dark matter (DM). The scattering of DM and quarks, mediated by three CP-even scalars in this model, is suppressed when masses of CP-even scalars are degenerate; that is called the ``degenerate scalar scenario''. 
Based on this scenario, we show that the strong first-order electroweak phase transition (EWPT) can be achieved without conflicting with constraints from the DM relic density and the direct detection experiments. We also discuss a shift of scalar trilinear coupling from the Standard Model prediction, which could be a test of this model in collider experiments.
\end{abstract}

\maketitle

\section{Introduction}\label{sec:intro}
Dark matter (DM), whose existence has been suggested by cosmological observations, predicts physics beyond the Standard Model (SM). Among the various DM candidates, weakly interacting massive particles (WIMP) are attractive because their abundance can be explained thermally. It is so-called the``WIMP miracle''. However, despite the vigorous searches, such as accelerator experiments and DM direct detection experiments have not found the DM signal. In particular, the LZ experiment~\cite{LZ:2024zvo} places very strong constraints on the scattering cross sections of WIMP-DM and nucleons, posing a major challenge for models that include WIMP-DM.

Possible scenarios consistent with the current direct detection experiments include
(1) DM is either sufficiently massive or has only tiny interactions with the SM particles
and (2) DM-nucleon scattering is suppressed by a built-in mechanism in the model. 
Here, we focus on the second scenario.
One model that realizes this scenario is the complex singlet scalar extension of the Standard Model (CxSM)~\cite{Barger:2008jx}, which includes a pseudo Nambu-Goldstone (pNG) DM.
In the CxSM, as the name suggests, the SM is extended by adding a complex scalar single particle field $S$. The imaginary part of $S$ behaves as the pNG DM, whose stability is guaranteed by the CP symmetry of the scalar potential. On the other hand, the real part of $S$ mixes with the SM Higgs boson to form the mass eigenstates $h_1$ and $h_2$. These two particles mediate DM-quark scattering. 
In ref.~\cite{Gross:2017dan}, the DM-quark scattering is suppressed due to the small momentum transfer.
However, this model has a so-called domain wall problem because the $Z_2$ symmetry ($S \to -S$) of the scalar potential $V$ is spontaneously broken due to the development of the vacuum expectation value (VEV) of the singlet scalar $(S)$. Therefore, in ref.~\cite{Abe:2021nih}, a linear term of $S$ ($V\supset a_1 S$) that breaks the $Z_2$ symmetry is introduced, and in this case,
it is proposed that when the masses of the two Higgs bosons are degenerate, i.e., $m_{h_1}\simeq m_{h_2}$, this scattering is suppressed. This suppression mechanism is called the degenerate scalar scenario.

On the other hand, baryon asymmetry of the Universe (BAU), along with DM, is one of the unsolved problems. The most testable scenario to realize BAU is electroweak baryogenesis (EWBG)~\cite{Kuzmin:1985mm, Rubakov:1996vz, Funakubo:1996dw, Riotto:1998bt, Trodden:1998ym, Bernreuther:2002uj, Cline:2006ts, Morrissey:2012db, Konstandin:2013caa, Senaha:2020mop} associated with the Higgs physics. EWBG requires strong first-order electroweak phase transition (EWPT), which suggests the need for an extension of the SM~\cite{Kajantie:1996mn, Rummukainen:1998as, Csikor:1998eu, Aoki:1999fi}. The CxSM is an attractive model for both DM and BAU, but in ref.~\cite{Cho:2021itv}, the poor compatibility between the degenerate scalar scenario and strong first-order EWPT.
In almost all models, strong first-order EWPT is caused by the effects of the finite temperature and the quantum corrections on the scalar potential, the latter is known as the Coleman-Weinberg potential. However, in some models with extended Higgs sector, the potential barrier is enhanced by mixing between the electroweak Higgs sector and a new scalar sector at the tree level. As investigated in ref.~\cite{Cho:2021itv}, the contribution of the latter is large in the CxSM, and a large $\delta_2$, which is the mixing parameter between the SM Higgs and the singlet scalar $S$, is favored to realize strong first-order EWPT. On the other hand, a small $\delta_2$ achieved through the degenerate scalar scenario suppresses the DM-quark scattering, so these two give contradictory conditions for the parameters. To explain them simultaneously, the DM mass needs to be half the Higgs mass, but here, the DM relic abundance is very small due to the resonance effect of the DM annihilation process and is far from the observed value $\Omega_\mathrm{DM}h^2 = 0.1200 \pm 0.0012$~\cite{Planck:2018vyg}.

In this paper, we consider a model in which an SU(2)$_L$ doublet scalar and a complex singlet scalar are added to the SM. Since this model is the two-Higgs doublet model (2HDM) with the complex singlet $ S $, we refer to this model as 2HDMS. In the 2HDMS, the CP-even scalars of the two Higgs doublet and the real part of $S$ mix to form three Higgs particles $H_{1,2,3}$. The imaginary part of $S$ acts as the DM. 
There are studies on 2HDMS, such as \cite{Jiang:2019soj,Zhang:2021alu, Biekotter:2021ovi,Biekotter:2022bxp}, but in these models, $S$ has the $Z_2$ symmetry and may suffer from the domain wall problem, so here we introduce a linear term for $S$ that breaks the symmetry. Furthermore, since $v_S$ becomes nonzero for any temperature due to $a_1\neq 0$, the EWPT is also different from the model with $a_1= 0$.
As with the CxSM mentioned earlier, we investigate whether the DM-quark scattering is suppressed when the mediator's masses are degenerate ($m_{H_1}\simeq m_{H_2}\simeq m_{H_3}$), in other words, we investigate the feasibility of the degenerate scalar scenario. We take into account four types of Yukawa interaction between the SM fermions and the SU(2)$_L$ doublet Higgs bosons, which are distinguished by the $Z_2$ charge of each fermion.
In the 2HDMS, the one-loop effect is most important for strong first-order EWPT, unlike the CxSM, where the tree-level structure is extremely important. As a result, it is shown that the degenerate scalar scenario is sufficiently valid even in the parameter region where strong first-order EWPT is achieved. In order to resolve the conflict between the degenerate scalar scenario and strong first-order EWPT, a two-component DM model that adds a fermion DM to the CxSM is also being studied~\cite{Cho:2023oad}, but here we explain the current DM abundance using only the pNG DM of the 2HDMS.

The structure of this paper is as follows. Sec.~\ref{sec:model} gives a brief review of the 2HDMS. After introducing the degenerate scalar scenario in Sec.~\ref{sec:dege}, to discuss the compatibility of the degenerate scalar scenario and strong first-order EWPT, we qualitatively analyze EWPT in Sec.~\ref{sec:EWPT}. Then, we numerically investigate the parameter region consistent from the perspective of DM and EWPT in Sec.~\ref{sec:nume}. Lastly, Sec.~\ref{sec:sum} summarizes this study.


\section{The Model}\label{sec:model} 
In the 2HDMS, the scalar potential is given as
\begin{align}
V_0=V_{0,\mathrm{2HDM}}+V_{0,\mathrm{S}}.
\label{pot}
\end{align}
The first term in r.h.s. in \eqref{pot} contains only the doublet Higgs fields $\Phi_1$ and $\Phi_2$
\begin{align}
V_{0,2 \mathrm{HDM}}\left(\Phi_1, \Phi_2\right)= & m_1^2 \Phi_1^{\dagger} \Phi_1+m_2^2 \Phi_2^{\dagger} \Phi_2-\left(m_3^2 \Phi_1^{\dagger} \Phi_2+\text { h.c. }\right) \nonumber \\
& +\frac{\lambda_1}{2}\left(\Phi_1^{\dagger} \Phi_1\right)^2+\frac{\lambda_2}{2}\left(\Phi_2^{\dagger} \Phi_2\right)^2+\lambda_3\left(\Phi_1^{\dagger} \Phi_1\right)\left(\Phi_2^{\dagger} \Phi_2\right)  \nonumber \\
& +\lambda_4\left(\Phi_1^{\dagger} \Phi_2\right)\left(\Phi_2^{\dagger} \Phi_1\right)+\left[\frac{\lambda_5}{2}\left(\Phi_1^{\dagger} \Phi_2\right)^2+\text { h.c. }\right],
\end{align}
where only $m_3^2$ term breaks the $Z_2$ symmetry of the doublets 
($\Phi_1 \to +\Phi_1,~\Phi_2 \to -\Phi_2$)
softly. 
The second term in \eqref{pot} is given by the singlet $S$ and doublets as
\begin{align}
V_{0,\mathrm{S}}\left(\Phi_1, \Phi_2, S\right)&=\frac{\delta_1}{2} \Phi_1^{\dagger} \Phi_1 |S|^2 + \frac{\delta_2}{2} \Phi_2^{\dagger} \Phi_2 |S|^2 + \frac{b_2}{2} |S|^2 + \frac{d_2}{4} |S|^4 \nonumber \\
&+ \left( a_1 S + \frac{b_1}{4} S^2 +\text { h.c.}\right),
\end{align}
where the terms in the first line are invariant under a global U(1) transformation 
$\qty(S \to e^{i\theta}S)$, and terms in the second line break the symmetry softly. 
In the following, the scalar potential is assumed to be CP-invariant, i.e., all coefficients are real.
Three scalar fields $\Phi_i~(i=1,2)$ and $S$ can be written as
\begin{align}
\Phi_i=\binom{\phi_i^{+}}{\frac{1}{\sqrt{2}}\left(v_i+h_i+i \eta_i\right)}, \quad S=\frac{1}{\sqrt{2}}\left(v_S+s+i \chi\right).
\end{align}
At this time, $v=\sqrt{v_1^2+v_2^2}$ = 246.22 GeV and $\tan\beta=v_2/v_1$ is defined.
The following
scalar particles are present in this model; the charged scalar $\qty(\phi_i^+)$, the CP-odd scalar $\qty(\eta_i)$, and the CP-even scalars $\qty(h_1,h_2,s)$. On the other hand, a pseudoscalar $\chi$ is the physical DM candidate.

First derivatives of $V_0$ with respect to $h_1,h_2$ and $s$ are respectively given by
\begin{align}
\left\langle\frac{\partial V_0}{\partial h_1}\right\rangle &=m_1^2v_1-m_3^2v_2+\frac{\lambda_1}{2} v_1^3+ \frac{\lambda_{345}}{2}v_1v_2^2+\frac{\delta_1}{4}v_1 v_S^2=0,\label{tad1} \\
\left\langle\frac{\partial V_0}{\partial h_2}\right\rangle &=m_2^2v_2-m_3^2v_1+\frac{\lambda_2}{2} v_2^3+\frac{\lambda_{345}}{2}v_1^2v_2+\frac{\delta_2}{4}v_2 v_S^2=0,\label{tad2} \\
\left\langle\frac{\partial V_0}{\partial s}\right\rangle & =\sqrt{2}a_1+\frac{b_1+b_2}{2}v_S+\frac{\delta_1}{4}v_1^2v_S+\frac{\delta_2}{4}v_2^2v_S+\frac{d_2}{4}v_S^3=0, \label{tad3}
\end{align}
where $\lambda_{345}=\lambda_3+\lambda_4+\lambda_5$.
The tree-level masses of the CP-even scalars are obtained by
\begin{align}
-\mathcal{L}_\text{mass}
& =
 \frac{1}{2}\left(\begin{array}{lll}
h_1 & h_2 & s
\end{array}\right) \mathcal{M}_{S}^{2}\left(\begin{array}{c}
h_1 \\
h_2 \\
s
\end{array}\right)
=
\frac{1}{2}\left(\begin{array}{lll}
H_1 & H_2 & H_3
\end{array}\right) 
O^\top \mathcal{M}_{S}^{2} O
\left(\begin{array}{c}
H_1 \\
H_2 \\
H_3
\end{array}\right)\nonumber \\
&= \frac{1}{2}\sum_{i=1}^3m_{H_i}^2H_i^2,
\label{mixingrelation}
\end{align}
with
\begin{align}
 \mathcal{M}_S^2=\left(\begin{array}{ccc}
m_3^2 \frac{v_2}{v_1}+\lambda_1 v_1^2 & -m_3^2+\lambda_{345} v_1 v_2 & \frac{\delta_1}{2} v_1 v_S \\
-m_3^2+\lambda_{345} v_1 v_2 & m_3^2 \frac{v_1}{v_2}+\lambda_2 v_2^2 & \frac{\delta_2}{2} v_2 v_S \\
\frac{\delta_1}{2} v_1 v_S & \frac{\delta_2}{2} v_2 v_S & -\frac{\sqrt{2} a_1}{v_S}+\frac{d_2}{2} v_S^2
\end{array}\right),
\end{align}
The mixing matrix $O$ is parametrized as
\begin{align}
O(\alpha_i)
&=\left(\begin{array}{ccc}
1 & 0 & 0 \\
0 & c_{3} & -s_{3} \\
0 & s_{3} & c_{3}
\end{array}\right)\left(\begin{array}{ccc}
c_{2} & 0 & -s_{2} \\
0 & 1 & 0 \\
s_{2} & 0 & c_{2}
\end{array}\right)\left(\begin{array}{ccc}
c_{1} & -s_{1} & 0 \\
s_{1} & c_{1} & 0 \\
0 & 0 & 1
\end{array}\right),
\end{align}
where $s_i=\sin\alpha_i$ and $c_i=\cos\alpha_i~(i=1,2,3)$. Note that the mixing matrix is orthogonal, i.e., 
\begin{align}
\sum_k O_{ik}O_{jk}=\delta_{ij}.
\label{orthgnl}
\end{align}
Three Higgs particles appear in the 2HDMS, and $H_1$ is the Higgs boson observed at the LHC experiment.

On the other hand, the charged Higgs $H^\pm$ and the CP-odd scalar $A$ are defined from $\phi^+_{1,2},~\eta_{1,2}$ as
\begin{align}
\quad\binom{\phi_1^{+}}{\phi_2^{+}}=R(\beta)\binom{G^{+}}{H^{+}},\quad\binom{\eta_1}{\eta_2}=R(\beta)\binom{G^0}{A},
\label{mssgn}
\end{align}
where
\begin{align}
R(\beta)=\left(\begin{array}{cc}
\cos \beta & -\sin \beta \\
\sin \beta & \cos \beta
\end{array}\right).
\end{align}
In \eqref{mssgn}, $G^{\pm}$ and $G^{0}$ denote Nambu-Goldstone bosons.
The mass eigenvalues of $H^\pm$ and $A$ are given by 
\begin{align}
m_{H^{ \pm}}^2&=\frac{m_3^2}{\sin \beta \cos \beta}-\frac{1}{2}\left(\lambda_4+\lambda_5\right) v^2, 
\label{mHch}\\
m_A^2&=\frac{m_3^2}{\sin \beta \cos \beta}-\lambda_5 v^2, 
\label{mA}
\end{align}
respectively.
Also, DM mass is given by
\begin{align}
m_{\chi}^2=-\frac{\sqrt{2}a_1}{v_S}-b_1.
\label{mDM}
\end{align}

Let us summarize our input parameters. There are 14 degrees of freedom in the scalar potential: $\{m_1^2,m_2^2,m_3^2,\lambda_1,\lambda_2,\lambda_3,\lambda_4,\lambda_5,\delta_1,\delta_2,b_2,d_2,a_1,b_1\}$. First, $\{m_3^2,a_1\}$ remain as input parameters. $\{m_1^2,m_2^2,b_2\}$ are determined by the tadpole conditions~\eqref{tad1}-\eqref{tad3} and $\{\lambda_4,\lambda_5,b_1\}$ are fixed by the particle masses~\eqref{mHch},\eqref{mA} and \eqref{mDM}. The other six Lagrangian parameters are determined by the mass matrix of the CP-even scalars. In Appendix~\ref{app:para}, we list relationships between the input parameters and original Lagrangian parameters. 

In order to prevent the tree-level FCNC in this model, we consider the case where one of the doublets couples with each fermion~\cite{Branco:2011iw}.
The Lagrangian of the Yukawa interaction is given as follows:
\begin{align}
-\mathcal{L}_{\text {Yukawa }}=\bar{Q}_L Y_u \tilde{\Phi}_u u_R+\bar{Q}_L Y_d \Phi_d d_R+\bar{L}_L Y_{\ell} \Phi_{\ell} \ell_R+ \mathrm{h.c.}, 
\end{align}
where $\Phi_f (f = u, d, \ell)$ is a Higgs doublet that couples to the fermion $f$
and $\tilde{\Phi}_u \equiv i\sigma_2 \Phi^*_u$ and $\sigma_2$ being the Pauli matrix.
${Q}_L$ and ${L}_L$ represent left-handed quarks and left-handed leptons, i.e., $Q_L =\left(u_L, d_L\right)^\top,~L_L =\left(\nu_L, e_L\right)^\top$. $u_R,~d_R$, and $\ell_R$ are right-handed up-type quarks, right-handed down-type quarks, and right-handed leptons. $Y_f (f = u, d, \ell)$ is the 3$\times$3 Yukawa matrix of the fermions. 
 As shown in Table~\ref{tab:Z2charge}, by assigning the $Z_2$ charge for the fermions, we can determine the combination of ${\Phi}_f~(f=u,d,\ell)$ and classify them into four types.

\begin{table}[t]
\center
\begin{tabular}{|c|c|c|c|c||c|c|c|}
\hline &  $Q_L, L_L$ & $u_R$ & $d_R$ & $\ell_R$  &  $\Phi_u$ & $\Phi_d$ & $\Phi_l$ \\
\hline Type-I &  $+$ & $-$ & $-$ & $-$ &   $\Phi_2$ & $\Phi_2$ & $\Phi_2$ \\
\hline Type-II & $+$ & $-$ & $+$ & $+$ &   $\Phi_2$ & $\Phi_1$ & $\Phi_1$ \\
\hline Type-X &  $+$ & $-$ & $-$ & $+$ &   $\Phi_2$ & $\Phi_2$ & $\Phi_1$ \\
\hline Type-Y & $+$ & $-$ & $+$ & $-$ &   $\Phi_2$ & $\Phi_1$ & $\Phi_2$ \\
\hline
\end{tabular}
\caption{Assignment of $Z_2$ charge to the fermions and combination of the doublets that couples to each fermion. 
The Higgs doublets $\Phi_1$ and $\Phi_2$ transform as $\Phi_1 \to +\Phi_1$ and $\Phi_2 \to -\Phi_2$, respectively. 
}
\label{tab:Z2charge}
\end{table}


\section{Degenerate scalar scenario}\label{sec:dege}
We study a suppression mechanism of the scattering process of DM $\chi$ off a quark $q$ 
\begin{align}
 \chi(p_1) + q (p_2) \to \chi(p_3) + q (p_4) 
\end{align}
in the 2HDMS. The scattering amplitude $\mathcal{M}$ is given by a sum of three amplitudes $\mathcal{M}_1, \mathcal{M}_2$ and $\mathcal{M}_3$ mediated by $H_1, H_2$ and $H_3$, respectively:
\begin{align}
 i\mathcal{M} 
&=
i\left(\mathcal{M}_1 + \mathcal{M}_2 +\mathcal{M}_3\right), \label{msumtr}
\\
 i\mathcal{M}_1 
&= 
-i 2 C_{\chi\chi H_1} C_{qq H_1}  \frac{1}{t-m_{H_1}^2} \bar{u}(p_4) u(p_2), \label{m1tr}
\\
 i\mathcal{M}_2 
&= 
-i 2 C_{\chi\chi H_2} C_{qq H_2}  \frac{1}{t-m_{H_2}^2} \bar{u}(p_4) u(p_2), \label{m2tr}
\\
 i\mathcal{M}_3 
&= 
-i 2 C_{\chi\chi H_3} C_{qq H_3}  \frac{1}{t-m_{H_3}^2} \bar{u}(p_4) u(p_2). \label{m3tr}
\end{align}
Here, $C_{S_1S_2S_3}$ denotes the scalar trilinear coupling for $S_1,S_2$ and $S_3$, i.e., $\mathcal{L}\supset C_{S_1S_2S_3}S_1S_2S_3$. Thus, $C_{\chi\chi H_i}$ is the three-point scalar coupling between two $\chi$ and $H_i$ ($i=1,2,3$). On the other hand, $C_{qq H_i}$ is the Yukawa coupling between the SM quarks $q$ and $H_i$, 
$\mathcal{L}\supset C_{q q H_i} \bar{q}_L q_R H_i$.
A variable $t$ describes a momentum transfer, $t\equiv \left(p_1 - p_3\right)^2$, and $u(p)~(\bar{u}(p))$ represents an incoming (outgoing) quark spinor with a momentum $p$. A factor 2 in the right-hand side of \eqref{m1tr}, \eqref{m2tr} and \eqref{m3tr} is a symmetry factor for the $\chi\chi H_i$ vertex.

First, we evaluate the degenerate scalar scenario in the mass eigenstates, and then use the gauge eigenstate to understand this scenario better.
In the 2HDMS, the scalar trilinear couplings are given as follows:
\begin{align}
C_{\chi\chi H_i}=\frac{1}{2v_S}\left(m_{H_i}^2+\frac{\sqrt{2}a_1}{v_S}\right)O_{3i},
\end{align}
which is derived from Eq.~\eqref{mixingrelation}.
On the other hand, the Yukawa interactions can be expressed as
\begin{align}
C_{qq H_i}=
\left\{\begin{array}{l}
\frac{m_q}{v_{1}}O_{1i} \quad\text{(When $q$ couples to $\Phi_1$)}, \\
\frac{m_q}{v_{2}}O_{2i} \quad\text{(When $q$ couples to $\Phi_2$)},
\end{array}\right.
\end{align}
where $m_q$ is the mass of quark $q$.
Therefore, the amplitude \eqref{msumtr} becomes
\begin{align}
i\mathcal{M}=
\left\{\begin{array}{l}
-i2 \frac{1}{2v_S}\sum_{i=1}^3\frac{m_{H_i}^2+\frac{\sqrt{2}a_1}{v_S}}{t-m_{H_i}^2} O_{3i}O_{1i}\bar{u}(p_4) u(p_2) \quad\text{(When $q$ couples to $\Phi_1$)}, \\
-i2 \frac{1}{2v_S}\sum_{i=1}^3\frac{m_{H_i}^2+\frac{\sqrt{2}a_1}{v_S}}{t-m_{H_i}^2} O_{3i}O_{2i}\bar{u}(p_4) u(p_2) \quad\text{(When $q$ couples to $\Phi_2$)}.
\end{array}\right.
\label{massdege}
\end{align}
If $a_1=0$, the amplitude vanishes in the low-energy limit ($t\to 0$) due to the orthogonality of the mixing matrix \eqref{orthgnl}. However, as mentioned above, in order to avoid the domain wall problem associated with the development of $v_S$, we are currently setting $a_1$ to a nonzero value. In this case, the Higgs masses $m_{H_i}$ must be degenerate to suppress the amplitude. This is the mechanism we call the degenerate scalar scenario.

Next, we discuss the degenerate scalar scenario in the gauge eigenstates to explore the origin of this mechanism.
The doublets that couple to each fermion differ depending on the types in the 2HDMS. Therefore, we investigate the suppression mechanism for (A) and (B): 
\begin{align*}
\mbox{
(A) }
 &
\mbox{Type-I, Type-X 
}
: 
\mbox{both up- and down-type quarks couple to $\Phi_2$}
\\
\mbox{
(B) 
}
&
\mbox{Type-II, Type-Y
}
: 
\mbox{up (down)-type quark couples to $\Phi_2$ ($\Phi_1$)}
\end{align*}

\subsection{Type-I, Type-X}\label{subsec:type1X} 
The Yukawa interaction for Type-I and Type-X is given by
\begin{align}
 -\mathcal{L}_{\mathrm{Yukawa}}
&=
\sum_{i=1}^3 C_{u u H_i} \overline{u_L} u_R H_i+\sum_{i=1}^3 C_{d d H_i} \overline{d_L} d_R H_i 
+ \textrm{h.c.}
,  
\end{align}
where  
\begin{align}
C_{u u H_i} = \frac{m_u}{v_2} O_{2 i},~C_{d d H_i} = \frac{m_d}{v_2} O_{2 i} ,
\end{align}
with $m_{u,d}$ being masses of up-type and down-type quarks. 
Here, the lepton sector is omitted since we focus only on the scattering with quarks.
On the other hand, the scalar trilinear interaction between $\chi$ and $H_i$ is described by 
\begin{align}
-\mathcal{L}
&\supset 
 C_{\chi\chi h_1} h_1 \chi^2 + C_{\chi\chi h_2} h_2 \chi^2 +  C_{\chi\chi s} s \chi^2
\nonumber \\
&=
C_{\chi\chi H_i} H_i \chi^2,
\end{align}
where 
\begin{align}
C_{\chi\chi H_i} &\equiv C_{\chi\chi h_1} O_{1i} + C_{\chi\chi h_2} O_{2i} +C_{\chi\chi s} O_{3i}. 
\label{ccssdd}
\end{align}
Explicit expression of these couplings are summarized in Appendix~\ref{app:bitri}.
Since the momentum transfer $t$ in the direct detection experiments is very small as compared to the mediator masses, i.e., $t\ll m_{H_i}^2$, 
the amplitudes related to up-type quarks and down-type quarks can each be written by
\begin{align}
 i \mathcal{M}_{\mathrm{up}}&=2 i \bar{u}\left(p_4\right) u\left(p_2\right)\frac{m_u}{v_2}\sum_{i=1}^3\frac{C_{\chi \chi H_i } O_{2 i}}{m_{H_i}^2 },\nonumber \\
 &=2 i \bar{u}\left(p_4\right) u\left(p_2\right)\frac{m_u}{v_2} \sum_{i=1}^3\left(C_{\chi \chi h_1} \frac{O_{1 i} O_{2 i}}{m_{H_i}^2}+C_{\chi \chi {h_2}} \frac{O_{2 i}^2}{m_{H_i}^2}+C_{\chi \chi s} \frac{O_{3 i} O_{2 i}}{m_{H_i}^2}\right), \label{cndtn1} \\
 i \mathcal{M}_{\mathrm{down}}&=2 i \bar{u}\left(p_4\right) u\left(p_2\right)\frac{m_d}{v_2}\sum_{i=1}^3\frac{C_{\chi \chi H_i } O_{2 i}}{m_{H_i}^2 },\nonumber \\
 &=2 i \bar{u}\left(p_4\right) u\left(p_2\right)\frac{m_d}{v_2} \sum_{i=1}^3\left(C_{\chi \chi h_1} \frac{O_{1 i} O_{2 i}}{m_{H_i}^2}+C_{\chi \chi {h_2}} \frac{O_{2 i}^2}{m_{H_i}^2}+C_{\chi \chi s} \frac{O_{3 i} O_{2 i}}{m_{H_i}^2}\right).
 \label{cndtn2}
\end{align}
Therefore, when three Higgs masses are degenerate, due to the orthogonality of the mixing matrix \eqref{orthgnl}, the terms that are proportional to $O_{1 i} O_{2 i}$ and $O_{3 i} O_{2 i}$ vanish. On the other hand, the term that is proportional to $O_{2 i}^2$ includes $C_{\chi \chi {h_2}}$. In this model, $C_{\chi \chi {h_2}}$ can be expressed as
\begin{align}
C_{\chi \chi {h_2}}= \frac{\delta_2}{4}v_2,
\end{align}
where
\begin{align}
\delta_2 &= \frac{2}{v_2 v_S} \sum_{i=1}^3  O_{2i} O_{3i} m_{H_i}^2 .
\label{del2}
\end{align}
$\delta_2$ is also suppressed by the orthogonality of the mixing matrix. Therefore, the amplitudes vanish in the degenerate scalar scenario $(m_{H_1}=m_{H_2}=m_{H_3})$. However, note that this suppression mechanism may not work well when $v_2$ and $v_S$, which appear in the denominator of Eq.~\eqref{del2}, are very small.

\begin{figure}[htpb]
\center
\includegraphics[width=15cm]{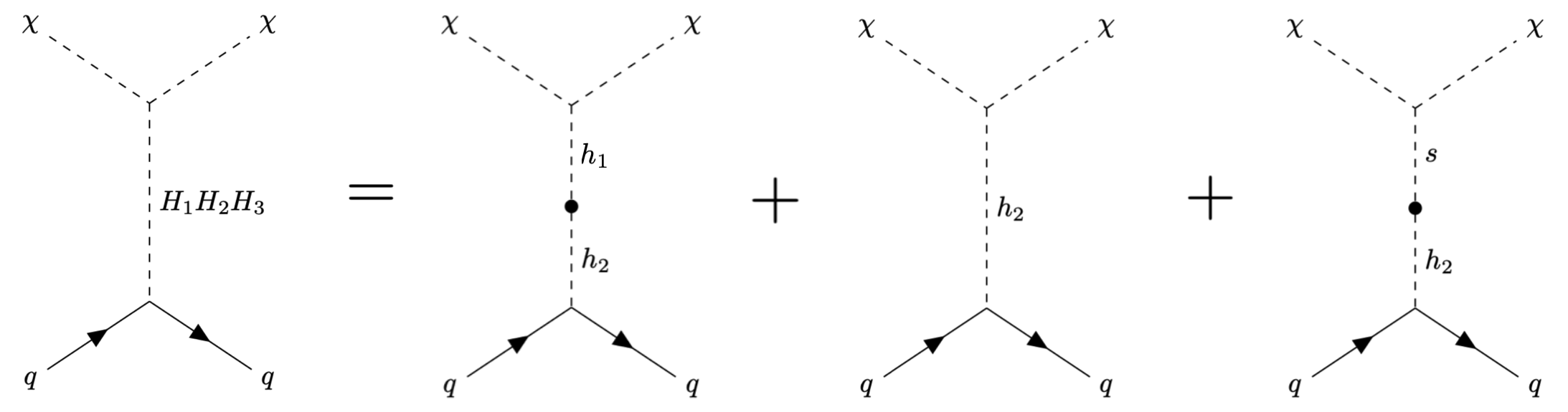}
\caption{Feynman diagrams of the DM-quark scattering in the mass and gauge eigenstates of scalar mediators. Both up-type quarks and down-type quarks couple to $\Phi_2$ in case A~(Type-I, Type-X).}
\label{fig:diagram1}
\end{figure}

We comment on the sum rule for the degenerate scalar scenario. The scattering of DM and quarks for case A~(Type-I, Type-X) is as shown in Fig.~\ref{fig:diagram1}. The left side uses the mass eigenstates, while the right side uses the gauge eigenstates. 
In order for the scalar $h_1$ and $s$ to couple with the quarks $q$, $h_1$ and $s$ must be converted into $h_2$ through the couplings $C_{h_1 h_2}$ and $C_{h_2 s}$ (the left and right diagram of the right-handed side in Fig.~\ref{fig:diagram1}). On the other hand, the amplitude mediated by $h_2$ is proportional to the coupling $C_{\chi\chi h_2}$ (the middle of the right-handed side diagram in Fig.~\ref{fig:diagram1}). 

The scalar potential in the 2HDMS allows us to rewrite the trilinear couplings $C_{\chi\chi h_1}, C_{\chi\chi h_2}$ and $C_{\chi\chi s}$ by the bilinear couplings $C_{h_1 s}, C_{h_2 s}$ and $C_{ss}$ as 
\begin{align}
 C_{\chi\chi h_1} &= \frac{A}{v_S} \left(C_{h_1 s} + \Delta_1\right), 
~~~
 C_{\chi\chi h_2} = \frac{A}{v_S} \left(C_{h_2 s} + \Delta_2\right), 
~~~
 C_{\chi\chi s} =
  \frac{A}{v_S} \left(C_{ss} + \Delta_3\right), 
\label{aaii}
\end{align}
where parameters $A, \Delta_1, \Delta_2$ and $\Delta_3$ are expressed by parameters besides $C_{h_1 s}, C_{h_2 s}$ and $C_{ss}$ in the scalar potential. The bilinear couplings are also given in Appendix~\ref{app:bitri}. Referring to Eqs.~\eqref{cndtn1}, \eqref{cndtn2} and extracting the part that is involved in suppressing the scattering, one gets
\begin{align}
\frac{C_{\chi\chi H_i} O_{2i}}{m_{H_i}^2}
&=
 \frac{A}{v_S}
\left\{
(C_{h_1 s}+ \Delta_1)O_{1i}+(C_{h_2 s}+ \Delta_2)O_{2i}+(C_{s s}+ \Delta_3)O_{3i}
\right\}\frac{O_{2i}}{m_{H_i}^2} ,\nonumber \\
&= \frac{A}{v_S
m_{H_i}^2}
\left
(O_{3i}O_{2i}m_{H_i}^2+\Delta_1 O_{1i}O_{2i}+\Delta_2 O_{2i}^2+\Delta_3 O_{3i}O_{2i}\right), 
\end{align}
where we use relations of coefficients between the gauge and mass eigenstates, i.e., $C_{h_1 s}O_{1i}+C_{h_2 s}O_{2i}+C_{s s}O_{3i}=O_{3i} m_{H_i}^2$.\footnote{This formula is derived from $O^\top M_S^2 O=\sum_{i=1}^3 m_{H_i}^2$ related to Eq.~\eqref{mixingrelation}.}
Therefore, the amplitude in the degenerate Higgs region vanishes when
\begin{align}
\Delta_2=0.
\label{cnd}
\end{align}

We show that the condition \eqref{cnd} holds for the scalar potential $V_0$ \eqref{pot}. 
In the 2HDMS, the couplings $C_{h_2 s}$ and $C_{\chi\chi h_2}$ are related to 
\begin{align}
V_0 &\supset\frac{\delta_2}{2} \Phi_2^{\dagger} \Phi_2 |S|^2  \nonumber \\
&\supset 
C_{\chi\chi h_2} h_2 \chi^2 + C_{h_2 s} h_2 s,
\end{align}
where
\begin{align}
C_{\chi\chi h_2} = \frac{\delta_2}{4}v_2, \quad
C_{h_2 s} = \frac{\delta_2}{2} v_2  v_S. 
\label{c_hs}
\end{align}
Comparing \eqref{c_hs} with \eqref{aaii}, we find 
\begin{align}
A &=\frac{1}{2}, 
~~~
\Delta_2 = 0. 
\label{cndtnmnml}
\end{align}
Therefore, in the degenerate limit of the mediator masses, the DM-quark scattering is suppressed as long as the potential \eqref{pot} is used. In other words, the mixing term between $\Phi_2$ and $S$ is constrained by the degenerate scalar scenario. 
For example, suppose that the mixing terms $\frac{c_2}{4} \Phi_2^{\dagger} \Phi_2 S$ is added to the potential \eqref{pot}. The couplings $C_{\chi\chi h_2}$ and $C_{h_2 s}$ become
\begin{align}
C_{\chi\chi h_2} = \frac{\delta_2}{4}v_2, \quad C_{h_2 s} = \frac{\delta_2}{2} v_2 v_S
+ \frac{\sqrt{2}c_2}{4}v_2. 
\end{align}
Then, 
\begin{align}
A &=\frac{1}{2}, 
~~~
\Delta_2 =-\frac{\sqrt{2}c_2}{4}v_2,
\end{align}
which is inconsistent with the condition \eqref{cnd}. Thus, if the mixing terms such as $\Phi_2^{\dagger} \Phi_2 S, \Phi_2^{\dagger} \Phi_2 S^2$ are added, the suppression mechanism does not work.\footnote{A more detailed analysis, including one-loop scattering, is given in Ref.~\cite{Cho:2023hek}. This discussion can be applied almost directly to the case of the 2HDMS.}

\subsection{Type-II, Type-Y}\label{subsec:type2Y}
The Yukawa interaction for Type-II and Type-Y is given by
\begin{align}
 -\mathcal{L}_{\mathrm{Yukawa}}
&=
\sum_{i=1}^3 C_{u u H_i} \overline{u_L} u_R H_i+\sum_{i=1}^3 C_{d d H_i} \overline{d_L} d_R H_i
+ \mathrm{h.c.}
,  
\end{align}
where  
\begin{align}
C_{u u H_i} = \frac{m_u}{v_2} O_{2 i},~C_{d d H_i} = \frac{m_d}{v_1} O_{1 i} .
\end{align}
The amplitude becomes
\begin{align}
i \mathcal{M}_\mathrm{up} & =2 i \bar{u}\left(p_4\right) u\left(p_2\right) \frac{m_u}{v_2} \sum_{i=1}^3\frac{C_{\chi \chi H_i} O_{2 i}}{m_{H_i}^2}\nonumber \\
&=2 i \bar{u}\left(p_4\right) u\left(p_2\right)\frac{m_u}{v_2}\sum_{i=1}^3\left(C_{\chi \chi h_1} \frac{O_{1 i} O_{2 i}}{m_{H_i}^2}+C_{\chi \chi {h_2}} \frac{O_{2 i}^2}{m_{H_i}^2}+C_{\chi \chi s} \frac{O_{3 i} O_{2 i}}{m_{H_i}^2}\right),\label{cndtn3} \\
i \mathcal{M}_\mathrm{down} & =2 i \bar{u}\left(p_4\right) u\left(p_2\right) \frac{m_d}{v_1}\sum_{i=1}^3 \frac{C_{\chi \chi H_i} O_{1 i}}{m_{H_i}^2}\nonumber \\
&=2 i \bar{u}\left(p_4\right) u\left(p_2\right)\frac{m_d}{v_1}\sum_{i=1}^3\left(C_{\chi \chi h_1} \frac{O_{1 i}^2}{m_{H_i}^2}+C_{\chi \chi {h_2}} \frac{O_{2 i}O_{1 i}}{m_{H_i}^2}+C_{\chi \chi s} \frac{O_{3 i}O_{1 i}}{m_{H_i}^2}\right).
 \label{cndtn4}
\end{align}
As for case A~(Type-I, Type-X), this scattering is suppressed when the three Higgs masses are equal. 
In order to suppress the down-type quarks scattering, it is necessary to suppress $C_{\chi \chi h_1}$, which can be expressed as
\begin{align}
C_{\chi\chi h_1} &= \frac{\delta_1}{4}v_1,
\end{align}
with
\begin{align}
\delta_1 &= \frac{2}{v_1 v_S} \sum_{i=1}^3 O_{1i} O_{3i} m_{H_i}^2.
\end{align}
Thus, in the degenerate scalar scenario, it is important to suppress $\delta_2$ for the up-type scattering and $\delta_1$ for the down-type scattering.

\begin{figure}[htpb]
\center
\includegraphics[width=15cm]{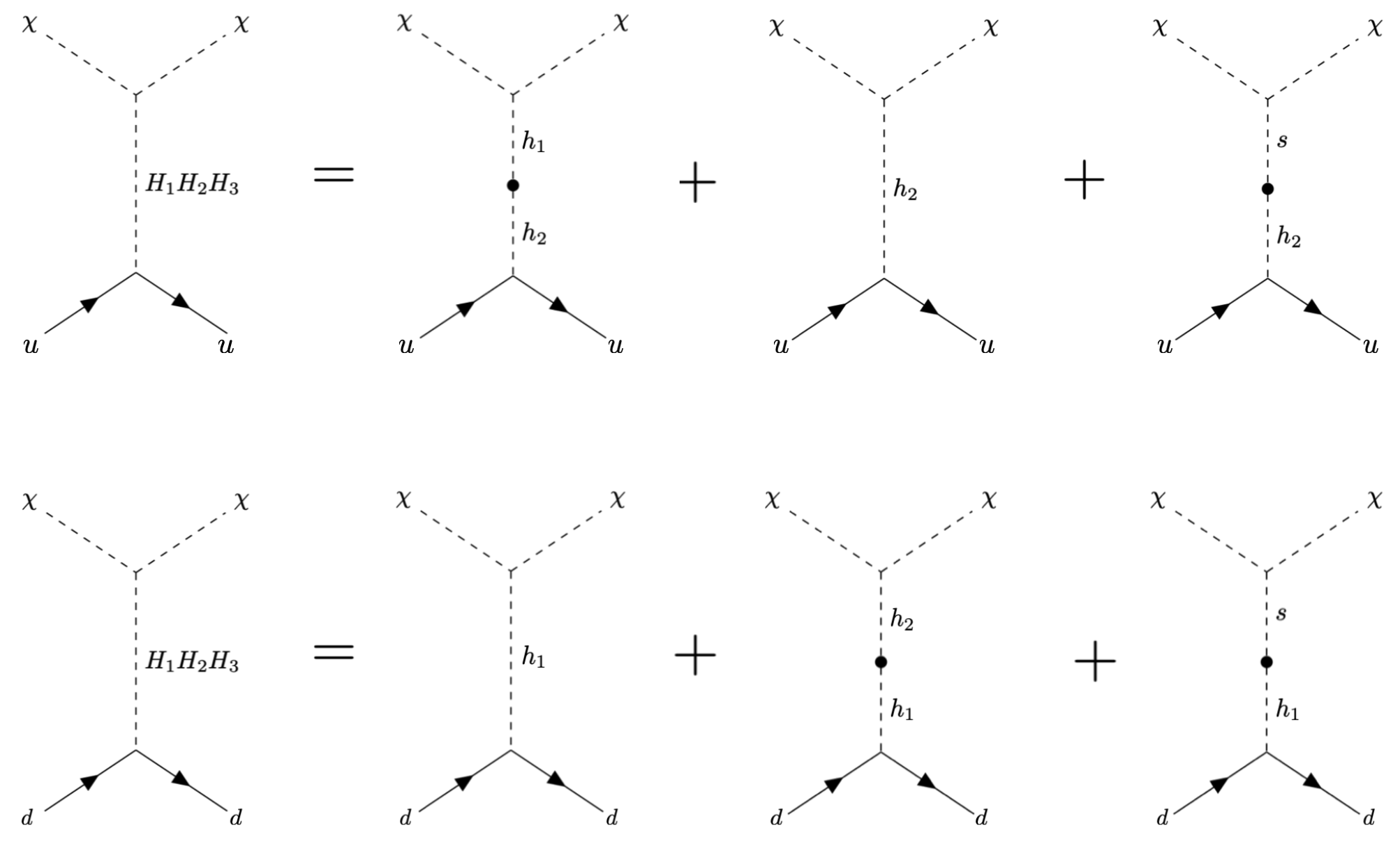}
\caption{Feynman diagrams of the DM-quark scattering in the mass and gauge eigenstates of scalar mediators. Up-type quarks $u$ couple to $\Phi_2$, while down-type quarks $d$ couple to $\Phi_1$ in case B~(Type-II, Type-Y).}
\label{fig:diagram2}
\end{figure}

We also mention the sum rule for the degenerate scalar scenario in this case. The scattering of DM and quarks for case B~(Type-II, Type-Y) is as shown in Fig.~\ref{fig:diagram2} and up-type quarks $u$ and down-type quarks $d$ couple to the different Higgs doublet.
The scattering of up-type quarks is the same as the discussion of case A, so we focus on those of down-type quarks. As can be seen from Eq.~\eqref{cndtn4}, the relevant part of the suppression mechanism is
\begin{align}
\frac{C_{\chi\chi H_i} O_{1i}}{m_{H_i}^2}
&=
 \frac{A}{v_S}
\left\{
(C_{h_1 s}+ \Delta_1)O_{1i}+(C_{h_2 s}+ \Delta_2)O_{2i}+(C_{s s}+ \Delta_3)O_{3i}
\right\}\frac{O_{1i}}{m_{H_i}^2} ,\nonumber \\
&= \frac{A}{v_S m_{H_i}^2}
\left
(O_{3i}O_{1i}m_{H_i}^2+\Delta_1 O_{1i}^2+\Delta_2 O_{2i}O_{1i}+\Delta_3 O_{3i}O_{1i}\right).
\end{align}
In this case, the amplitude in the degenerate Higgs region vanishes when
\begin{align}
\Delta_1=0
\label{cnd2}
\end{align}

In the 2HDMS, the couplings $C_{h_1 s}$ and $C_{\chi\chi h_1}$ are related to 
\begin{align}
V_0 &\supset\frac{\delta_1}{2} \Phi_1^{\dagger} \Phi_1 |S|^2  \nonumber \\
&\supset 
C_{\chi\chi h_1} h_1 \chi^2 + C_{h_1 s} h_1 s,
\end{align}
where
\begin{align}
C_{\chi\chi h_1} = \frac{\delta_1}{4}v_1,\quad
C_{h_1 s} = \frac{\delta_1}{2} v_1 v_S. 
\label{c_h1s}
\end{align}
Comparing \eqref{c_h1s} with \eqref{aaii}, we find 
\begin{align}
A &=\frac{1}{2}, 
~~~
\Delta_1 = 0. 
\label{cndtnmnml2}
\end{align}
Thus, in the degenerate scalar scenario, the DM-quark scattering is suppressed with the potential \eqref{pot} same as case A.

\subsection{Higgs search}\label{subsec:Higgssearch}
We mention the usefulness of the degenerate scalar scenario in the Higgs search experiments at the LHC. 
In the 2HDMS, the couplings between the Higgs boson $H_i$ and the SM particle $X$ are given by the SM couplings multiplied by the mixing matrix elements $O_{1i}$. Taking the decay widths of $H_i$ as an example, we obtain
\begin{align}
& \Gamma_{H_1 \rightarrow XX}=\Gamma_{h \rightarrow XX}^\mathrm{SM}\left(m_{H_1}\right) \times O_{11}^2, \\
& \Gamma_{H_2 \rightarrow XX}=\Gamma_{h \rightarrow XX}^\mathrm{SM}\left(m_{H_2}\right) \times O_{12}^2, \\
& \Gamma_{H_3 \rightarrow XX}=\Gamma_{h \rightarrow XX}^\mathrm{SM}\left(m_{H_3}\right) \times O_{13}^2,
\end{align}
where $\Gamma_{h \rightarrow XX}^\mathrm{SM}\left(m_{H_i}\right)$ is the decay width of the SM Higgs $h$ as a function of $m_{H_i}$. When we add up all the decay widths with $m_{H_i} \simeq m_h=125~\mathrm{GeV}$,
\begin{align}
\Gamma_{H_1 \rightarrow XX}+\Gamma_{H_2 \rightarrow XX}+\Gamma_{H_3 \rightarrow XX}\simeq \Gamma_{h \rightarrow XX}^\mathrm{SM}\left(m_{h}\right)
\end{align}
is obtained. This SM-like process is effective for all mixing angles, and $H_i$ cannot be distinguished in the current experiment~\cite{Abe:2021nih}.

\section{Qualitative discussion for electroweak phase transition}\label{sec:EWPT}
We discuss the compatibility of the degenerate scalar scenario and EWPT qualitatively. 
As mentioned in Sec.~\ref{sec:dege}, the orthogonality of the mixing matrix \eqref{orthgnl} is important for the degenerate scalar scenario. Here, using the elements of the mixing matrix, the mixing parameters for SU(2)$_L$ doublet-singlet in the scalar potential can be expressed as
\begin{align}
\delta_1 &= \frac{2}{v_1 v_S} \left( O_{1i} O_{3i} m_{H_i}^2  \right), \\
\delta_2 &= \frac{2}{v_2 v_S} \left( O_{2i} O_{3i} m_{H_i}^2  \right).
\end{align}
As we discussed in Sec.~\ref{sec:dege}, it can be seen that the DM-quark scattering is controlled by $\delta_1$ and $\delta_2$. In other words, the degenerate scalar scenario is realized by $\delta_1$ and $\delta_2$.
Strong first-order EWPT is essential for EWBG, and the condition is~\cite{Arnold:1987mh,Bochkarev:1987wf,Funakubo:2009eg};
\begin{align}
\frac{v_C}{T_C} \gtrsim 1,\label{decoupling}
\end{align}
where $T_C$ is the critical temperature at which the effective potential has two degenerate minima, and $v_C = \sqrt{v_{1C}^2 + v_{2C}^2}$ is the nonzero Higgs VEV at $T=T_C$. Possible origins of first-order EWPT are the finite-temperature boson loop  
and the structure of the tree-level potential. In the former case, the addition of multiple bosons is effective, and the 2HDM specializes in this type of EWPT. On the other hand, in the latter case, the existence of a single scalar is important, and the CxSM causes this type of EWPT. Since the 2HDMS has multiple bosons and a singlet scalar that does not exist in the SM, EWPT of both types can occur. First, we qualitatively analyze tree-level driven EWPT using the so-called high temperature (HT) potential, which is composed of the tree-level potential and thermal masses.
In the high-temperature expansion of the finite temperature effective potential, we can obtain not only the thermal masses but also the thermal cubic terms. However, here, we only add the thermal masses to the tree-level potential since we want to check whether first-order EWPT occurs due to the effect of the doublet-singlet mixing of the tree-level. Generally, first-order EWPT would be enhanced even more by adding the thermal cubic terms.

EWPT pattern of 2HDMS is 
$\qty(\langle \Phi_1 \rangle, \langle \Phi_2 \rangle, \langle S \rangle)=(0,0,v_S') \to (v_1,v_2,v_S)$, 
where $\langle \Phi_i \rangle$ and $\langle S \rangle$ are defined as the VEVs of the doublet fields $\Phi_i$ and the singlet field $S$, respectively. To discuss EWPT,  we take the Landau gauge and assume that the charged scalar bosons do not have VEVs at any temperature in order to conserve the U(1)$_{\mathrm{QED}}$.
We denote the classical background fields of the Higgs doublets and singlet as
\begin{align}
\langle \Phi_i\rangle=\left(\begin{array}{ll}
0 & \varphi_i
\end{array}\right)^\top / \sqrt{2},\quad\langle S\rangle=\varphi_S / \sqrt{2}.
\end{align}
The tree-level potential \eqref{pot} can be rewritten as
\begin{align}
V_0\left(\varphi_1, \varphi_2 , \varphi_S\right)&=\frac{m_1^2}{2} \varphi_1^2+\frac{m_2^2}{2} \varphi_2^2-m_3^2 \varphi_1 \varphi_2+\frac{\lambda_1}{8} \varphi_1^4+\frac{\lambda_2}{8} \varphi_2^4+\frac{\lambda_{345}}{4}\left(\varphi_1 \varphi_2\right)^2 \nonumber \\
&+\frac{\delta_1}{8} \varphi_1^2\varphi_S^2+\frac{\delta_2}{8} \varphi_2^2 \varphi_S^2+\frac{b_2}{4}\varphi_S^2+\frac{d_2}{16}\varphi_S^4 +\sqrt{2}a_1\varphi_S+\frac{b_1}{4}\varphi_S^2 .
\end{align}
The HT potential is defined as
\begin{align}
V^{\mathrm{HT}}\left(\varphi_1,\varphi_2, \varphi_S ; T\right)=V_0\left(\varphi_1,\varphi_2, \varphi_S\right)+\left(\Sigma_{h1} \varphi_1^2+\Sigma_{h2} \varphi_2^2+\Sigma_S \varphi_S^2\right) T^2,
\label{HTpote}
\end{align}
where
\begin{align}
\Sigma_{h1}&=\frac{\delta_1}{48} + \frac{g_1^2}{32} + \frac{3 g_2^2}{32} + \frac{\lambda_1}{8} + \frac{\lambda_3}{12} + \frac{\lambda_4}{24} + \frac{y_t^2}{8} + \frac{y_b^2}{8}, \label{sigmah1}\\
\Sigma_{h2}&=\frac{\delta_2}{48} + \frac{g_1^2}{32} + \frac{3 g_2^2}{32} + \frac{\lambda_2}{8} + \frac{\lambda_3}{12} + \frac{\lambda_4}{24} + \frac{y_t^2}{8} + \frac{y_b^2}{8}, \label{sigmah2}\\
\Sigma_{S}&=\frac{d_2}{24} + \frac{\delta_1}{24} + \frac{\delta_2}{24},
\end{align}
with $g_2$, $g_1$, $y_t$ and $y_b$ being SU(2)$_L$ , U(1)$_Y$, top Yukawa and bottom Yukawa couplings.
It should be emphasized that the thermal cubic term of the field is necessary for loop-driven EWPT. Since the HT potential does not include that term, it is optimal for analyzing EWPT derived from the tree-level potential.

For convenience, $\varphi_i$ and $\varphi_S$ are parameterized as
\begin{align}
\varphi_1=z \cos \theta \cos \beta,\quad \varphi_2=z \cos \theta \sin \beta,\quad \varphi_S=z \sin \theta+{v}_S',
\end{align}
where $v_S'$ denotes the minimum on the $\varphi_S$ axis, which is always nonzero because of $a_1\neq 0$ in
our model. The HT potential \eqref{HTpote} becomes
\begin{align}
 &V^{\mathrm{HT}}(z, \theta, \beta ; T)=c_0+c_1 z+\left(c_2+c_2^{\prime} T^2\right) z^2-c_3 z^3+c_4 z^4.
\label{simplifyHT}
\end{align}
Eq.~\eqref{simplifyHT} is composed of the tree-level potential and the thermal masses, and there is no thermal cubic term like $-c'_3 T z^3$. This means that strong first-order EWPT could be achieved even if $c'_3=0$ in this model.
After shifting the vacuum energy with $c_0=0$, the above potential at $T=T_C$ takes the form
\begin{align}
V^{\mathrm{HT}}\left(z, \theta, \beta ; T_C\right)=c_4 z^2\left(z-z_C\right)^2, \quad z_C=\frac{c_3}{2 c_4},
\end{align}
where
\begin{align}
c_3&=\frac{1}{4} v_S' \sin \theta \left( \delta_1 \cos^2 \beta \cos^2 \theta + \delta_2 \cos^2 \theta \sin^2 \beta + d_2 \sin^2 \theta \right), \\
 c_4&=\frac{1}{16} \left( 2 \lambda_1 \cos^4 \beta \cos^4 \theta + 2 \lambda_2 \cos^4 \theta \sin^4 \beta + 2 \delta_2 \cos^2 \theta \sin^2 \beta \sin^2 \theta \right. \nonumber \\
&\left. + d_2 \sin^4 \theta + 2 \cos^2 \beta \left( 2 \lambda_{345} \cos^4 \theta \sin^2 \beta + \delta_1 \cos^2 \theta \sin^2 \theta \right) \right).
\end{align}
From this potential, $v_C$ and $T_C$ are derived, which are
\begin{align}
v_C&=\sqrt{\frac{{v}_{SC}' ({v}_{SC}'-{v}_{SC}) 
\qty( \delta_1  \cos^2 \beta_C + \delta_2  \sin^2 \beta_C ) }{\lambda_1  \cos_{\beta_C}^4 +  \lambda_2  \sin_{\beta_C}^4 + 2 \lambda_{345} \sin_{\beta_C}^2 \cos_{\beta_C}^2 }}\label{vC}
 \\
T_C&=\sqrt{\frac{-m_1^2\cos^2{\beta_C}-m_2^2\sin^2{\beta_C}+2m_3^2\cos{\beta_C} \sin{\beta_C}-\frac{1}{4}{v}_{SC}'(\delta_1 \cos^2{\beta_C}+\delta_2 \sin^2{\beta_C})}{\Sigma_{h1}\cos^2{\beta_C}+\Sigma_{h2}\sin^2{\beta_C}}}.
\label{TC}
\end{align}
The physical quantities evaluated by $T_C$ are labeled with the subscript $C$. We can find that the condition of strong first-order EWPT \eqref{decoupling} requires large $\delta_1$ and $\delta_2$. 
If, in Eq.~\eqref{TC},  the following relation holds 
\begin{align}
-m_1^2\cos^2{\beta_C}-m_2^2\sin^2{\beta_C}+2m_3^2\cos{\beta_C} \sin{\beta_C}=0, 
\end{align}
it seems that $v_C/T_C$ is independent from both $\delta_1$ and $\delta_2$ 
since the terms related to $\delta_1$ and $\delta_2$ are canceled out between $v_C$ and $T_C$. However, there are $\delta_1$ and $\delta_2$ in $\Sigma_{h1}$ \eqref{sigmah1} and $\Sigma_{h2}$ \eqref{sigmah2}, and large $\delta_1$ and $\delta_2$ are still preferred to satisfy the condition of strong first-order EWPT. 

However, as mentioned above, small $\delta_1,\delta_2$ are essential for realizing the degenerate scalar scenario. Therefore, we do not entrust EWPT to the tree-level double-single mixing $\delta_1,\delta_2$, but to the thermal loop. The 1-loop effective potential can be written as follows:
\begin{align}
V_{\text{eff}}\left(\varphi_1, \varphi_2 , \varphi_S ; T\right)&=V_{0}(\varphi_1, \varphi_2 , \varphi_S)+V_{1}(\varphi_1, \varphi_2 , \varphi_S;T), \nonumber \\
&=V_{0}(\varphi_1, \varphi_2 , \varphi_S)+\sum_i n_i\left[V_{\mathrm{CW}}\left(\bar{m}_i^2\right)+\frac{T^4}{2 \pi^2} I_{B, F}\left(\frac{\bar{m}_i^2}{T^2}\right)\right], \label{eff}
\end{align}
where $T$ represents temperature. The subscript $i$ represents the particles contained in the model, and $n_i$ is the degree of freedom of each particle. The effective potentials at zero temperature and at the finite temperature are given by~\cite{Weinberg:1973am,Jackiw:1974cv,Dolan:1973qd}
\begin{align}
V_{\mathrm{CW}}\left(\bar{m}_i^2\right) &=\frac{\bar{m}_i^4}{64 \pi^2}\left(\ln \frac{\bar{m}_i^2}{\bar{\mu}^2}-c_i\right), \\
I_{B, F}\left(a^2\right) &=\int_0^{\infty} d x x^2 \ln \left(1 \mp e^{-\sqrt{x^2+a^2}}\right) \label{finite},
\end{align}
where $c_i=3/2$ for scalars and fermions and $c_i=5/6$ for gauge bosons. $\bar{m_i}$ is the field-dependent masses of each particle $i$.  $I_B$ with the minus sign represents the boson contribution, while $I_F$ with the plus sign represents the fermion one.  $\bar{\mu}$ is a renormalization scale. 

In the following calculations, the condition \eqref{decoupling} is checked using the 1-loop effective potential \eqref{eff}. In order to prevent the breakdown of perturbation theory due to boson multi-loop, the Parwani resummation, in which all the Matsubara frequency modes are resummed, is used~\cite{Parwani:1991gq}. Specifically, field-dependent masses $\bar{m}_i^2$ in Eq.~\eqref{eff} are replaced by thermally corrected ones. 

\section{Numerical results}\label{sec:nume}
Before giving benchmark points, we mention the theoretical and experimental constraints imposed on the parameters. A requirement on the scalar potential that is bounded from below is given by~\cite{Nie:1998yn,Kanemura:1999xf}
\begin{align}
\lambda_1>0,~\lambda_2>0,~\lambda_3+\lambda_4-\lambda_5>-\sqrt{\lambda_1 \lambda_2},~d_2>0.
\end{align}
The quartic couplings should also satisfy the following conditions from the tree-level unitarity~\cite{Chen:2014ask,Akeroyd:2000wc}
\begin{align}
\lambda_1<\frac{8\pi}{3},~\lambda_2<\frac{8\pi}{3},~\lambda_{345}<{8\pi},~\delta_1<16\pi,~\delta_2<16\pi,~d_2<\frac{16\pi}{3}.
\end{align}
The constraint from the perturbativity is as follows~\cite{Aoki:2021oez}:
\begin{align}
\left|\lambda_j\right|,d_2<4 \pi~(j=1-5).
\end{align}
Furthermore, the sizes of $\delta_1$ and $\delta_2$ could be constrained by a global minimum condition at zero temperature~\cite{Chen:2020wvu}. The energy of the electroweak vacuum has to be lower than that of the local vacuum on the $\varphi_S$ axis.
Experimental constraints here are primarily related to electroweak precision data and flavor experiments.
By assuming a mass degeneracy between the charged scalar boson $H^\pm$ and the neutral scalar boson $H$ or the CP-odd scalar boson $A$, i.e., $m_{H^{ \pm}} \simeq m_H \text { or } m_A$, the electroweak precision data is satisfied~\cite{Haber:2010bw}.
In addition, for Type-I 2HDMS, $B_d \rightarrow \mu \mu$ process requires $\tan{\beta}\gtrsim 1.75$ for $m_{H^{\pm}}=500$ GeV~\cite{Haller:2018nnx}. We should emphasize that the Higgs coupling measurement usually places a constraint on $\cos(\beta-\alpha)$, but not now since our analysis is based on the degenerate scalar scenario. 

\begin{table}[t]
\center
\begin{tabular}{|c|c|c|c|c|c|c|c|c|c|c|c|c|c|c|}
\hline
Inputs & $v$ & $v_S$ & $m_{H_1}$ & $m_{H_2}$ & $m_{H_3}$ & $m_{H^\pm}$ & $m_{A}$ & $m_{\chi}$ & $\alpha_1$ & $\alpha_2$  & $\alpha_3$  & $\tan{\beta}$ & $a_1$  &~$m_3^2$~  \\ \hline
BP1 & 246.22 & 246.22 & 125.0 & 124.5 & 124.0 & 500 & 500 & 140.19 & $\pi/4$ & $\pi/4$ & 0.01 & 2.0 & $-$6550 & 10 \\ \hline
BP2 & 246.22 & 200 & 125.0 & 124.5 & 124.0 & 500 & 500 & 130.85 & $\pi/4$ & $\pi/4$ & 0.01 & 2.0 & $-$700000 & 10 \\ \hline
BP3 & 246.22 & 30 & 125.0 & 124.5 & 124.0 & 500 & 500 & 122.60 & $\pi/4$ & $\pi/4$ & 0.01 & 2.0 & $-$323000 & 10 \\ \hline \hline
Outputs & $\delta_1$ & $\delta_2$ & $d_2$ & $\lambda_1$ & $\lambda_2$ & $\lambda_3$ & $\lambda_4$ & $\lambda_5$ & $m_1^2$ & $m_2^2$ & $b_1$ & $b_2$ & $v_1$ & $v_2$ \\ \hline
BP1 & 0.0069 & 0.0017 & 0.51 & 1.27 & 0.32 & 8.25 & $-$4.12 & $-$4.12 & $-$7863 & $-$7812 & $-$19616 & 4176 &110 &220 \\ \hline
BP2 & 0.0085 & 0.0020 & 0.53 & 1.27 & 0.32 & 8.25 & $-$4.12 & $-$4.12 & $-$7843 & $-$7807 & $-$12172 & 11450 &110 &220 \\ \hline
BP3 & 0.057 & 0.014 & 0.54 & 1.27 & 0.32 & 8.25 & $-$4.12 & $-$4.12 & $-$7771 & $-$7790 & 195.6 & 29339 &110 &220 \\ \hline
\end{tabular}
\caption{Input and output parameters. These BPs reproduce the observed DM relic density \eqref{relicobserve}.}
\label{tab:BP}
\end{table}

The input and output parameters that satisfy the above constraints are summarized in Table~\ref{tab:BP}. The masses of three Higgs bosons are degenerate within 1 GeV. In addition, these three benchmark points (BPs) reflect the EWPT discussion in Sec.~\ref{sec:EWPT}. The difference between input parameters among BPs is the size of $v_S$. The parameter $a_1$ is changed to keep the value of $d_2$ perturbatively appropriate.
The observed DM relic density is known as~\cite{Planck:2018vyg}
\begin{align}
\Omega_\mathrm{DM} h^2=0.1200 \pm 0.0012,
\label{relicobserve}
\end{align}
and we choose the DM mass that reproduces this value at each BP.

\begin{table}[t]
\center
\begin{tabular}{|c|c|c|c|c|c|c|c|c|}
\hline
 & $v_{1C}$ [GeV] & $v_{2C}$ [GeV] & $v_{SC}$ [GeV] & $v_{SC}^{\prime}$ [GeV]& $v_C/T_C$ \\ \hline
~BP1~ & 105.8 & 213.2 & 245.6 & 246.2 & $\frac{238.0}{69.6}=3.4$ \\ \hline
~BP2~ & 105.8 & 213.2 & 199.5 & 200.1 & $\frac{238.0}{69.6}=3.4$ \\ \hline
~BP3~ & 105.7 & 213.2 & 30.0 & 30.6 & $\frac{238.0}{69.6}=3.4$ \\ \hline
\end{tabular}
\caption{$T_C$ and corresponding VEVs for three BPs in Table.~\ref{tab:BP}. Here, the calculations are performed by \texttt{cosmoTransitions}~\cite{Wainwright:2011kj}.}
\label{tab:TCvC}
\end{table}

The public code \texttt{cosmoTransitions}~\cite{Wainwright:2011kj} is used to evaluate EWPT. In Table~\ref{tab:TCvC}, $T_C$ and corresponding VEVs for BPs are shown, and strong first-order EWPT occurs in all BPs. 
$v_C/T_C$ is determined roughly by the ratio of the cubic and quartic terms of the field. The bosons masses are important for the former, and $\lambda_i$ is important for the latter.
Since the boson masses are almost the same and the value of $\lambda_i$ is completely the same for all BPs, the Higgs VEVs and $T_C$ are almost identical in three cases.
In addition, the small change in the singlet VEV before and after EWPT shows that the tree-level effect by the singlet introduction does not cause EWPT this time. In fact, when using the HT potential \eqref{HTpote}, EWPT becomes second-order.

\begin{figure}[htpb]
\center
\includegraphics[width=8cm]{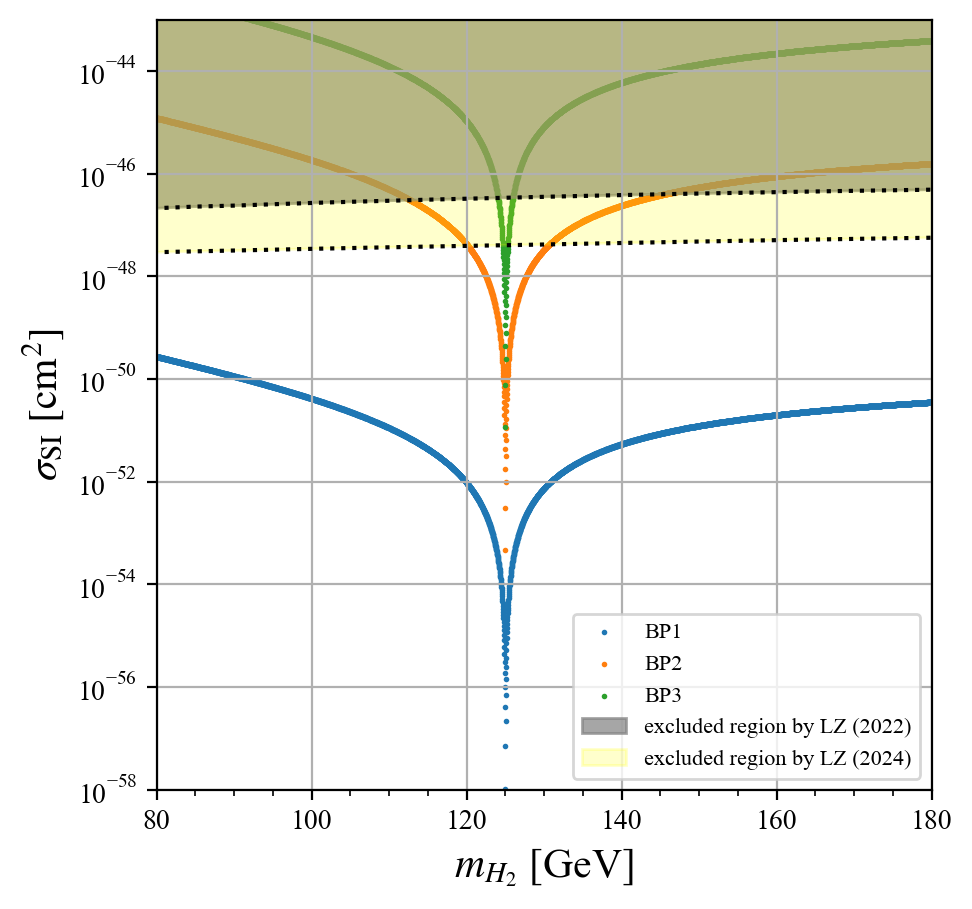}
\caption{Spin-independent cross section for the scattering between the DM and nucleus $\sigma_\mathrm{SI}$ as a function of $m_{H_2}$.}
\label{fig:dege}
\end{figure}

Next, we show the results of numerical calculations related to the DM using the public code \texttt{micrOMEGAs}~\cite{Belanger:2010pz,Belanger:2020gnr}. In Fig.~\ref{fig:dege}, the spin-independent cross section for the scattering between the DM and nucleus $\sigma_\mathrm{SI}$ as a function of $m_{H_2}$ is shown. Here, $m_{H_2}$ is used once as a variable, and the other parameters are BPs in Table~\ref{tab:BP}. The shaded regions are excluded by the LZ experiment. For all BPs, there is a large dip around $m_{H_2}\simeq 125$ GeV. This indicates that the degenerate scalar scenario is important for the suppression of DM and nucleon scattering. 
Although the value of $\delta_2$ varies between the BPs, $a_1$ may be more closely controlling the overall magnitude of $\sigma_\mathrm{SI}$. As can be seen from Eq.~\eqref{massdege}, $\sigma_\mathrm{SI}$ is proportional to $a_1^2$. For example, the difference in $\sigma_\mathrm{SI}$ between BP1 and BP2 is roughly $10^4$, which is due to the fact that $a_1^2\mathrm{(BP2)}/a_1^2\mathrm{(BP1)} \simeq 10^{4}$. $a_1$ is important for adjusting $d_2$ to an appropriate value, but it also affects the magnitude of $\sigma_\mathrm{SI}$ in this way. On the other hand, if the Higgs masses are degenerate, the scattering is suppressed regardless of the value of $a_1$. In other words, the degenerate scalar scenario is also valid when $a_1=0$, but to avoid the domain wall problem, we should consider $a_1\neq 0$.
Thus, Although the elastic neutrino-nucleon scattering makes DM detection difficult in some regions, this does not rule out the existence of the WIMP-DM~\cite{OHare:2021utq}.

\begin{figure}[t]
\center
\includegraphics[width=7cm]{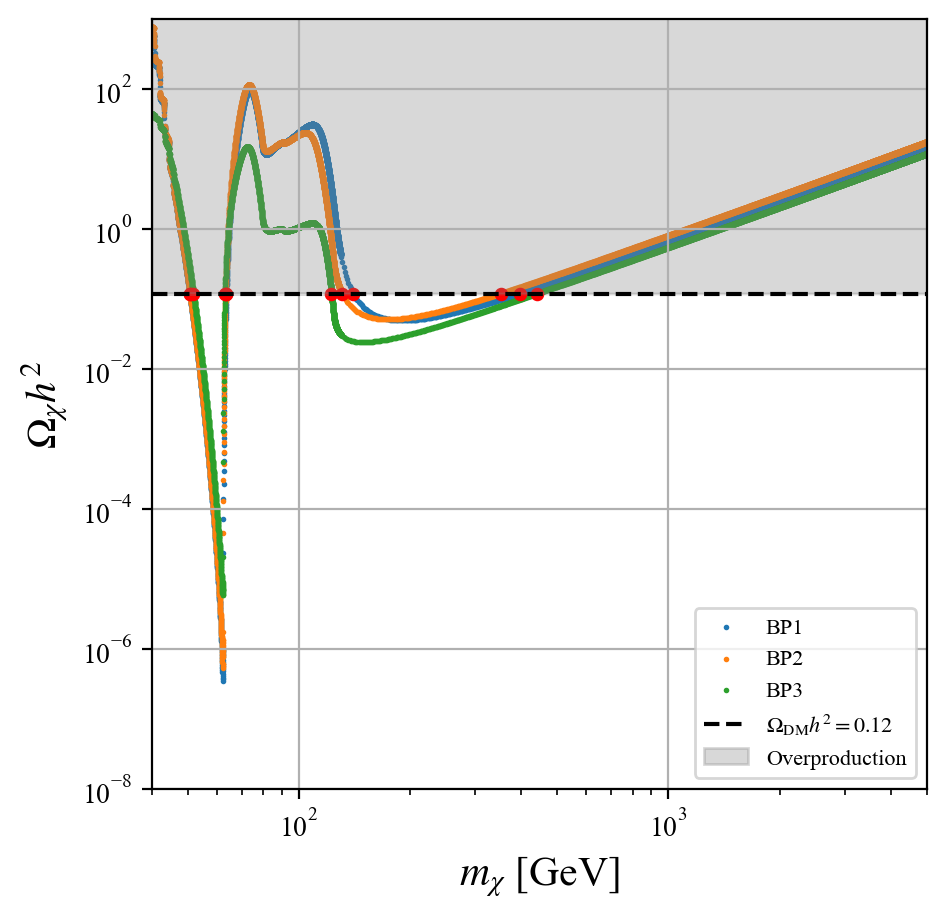}
\includegraphics[width=7cm]{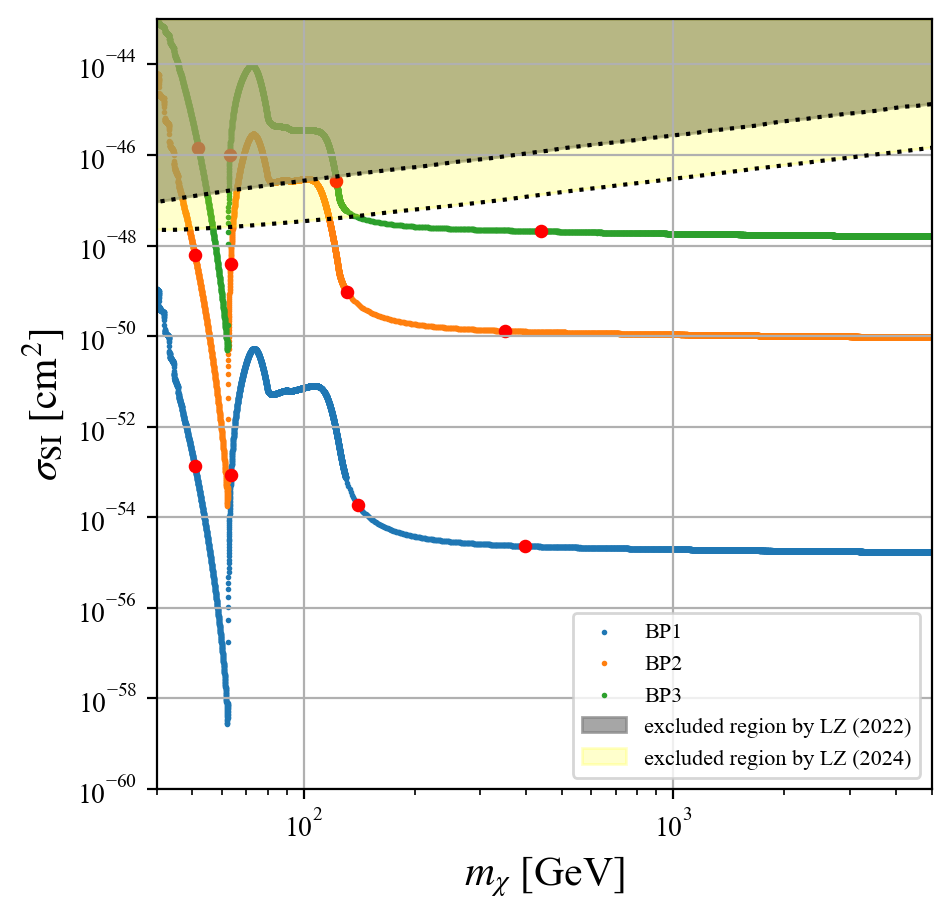}
\caption{(Left panel) DM relic density $\Omega_{\chi}h^2$ as a function of DM mass $m_chi$ for the BPs in Table.~\ref{tab:BP} is shown. The black dashed line represents the observed value \eqref{relicobserve}, and the region above this corresponds to overproduction. (Right panel) scaled scattering cross section between the DM $\chi$ and quarks $\tilde{\sigma}_{\mathrm{SI}}$ as a function of DM mass $m_\chi$ for all BPs is shown. The dotted line represents the bound of the LZ experiment, and the region above this is excluded. The red dots on both panels indicate where the observed relic density and the predicted one from the 2HDMS are equal.}
\label{fig:DM}
\end{figure}

The left panel of Fig.~\ref{fig:DM} shows $\Omega_{\chi}h^2$ as a function of $m_\chi$ for all BPs in Table.~\ref{tab:BP}. The value of the DM relic density $\Omega_\mathrm{DM} h^2$ must not exceed the observed value \eqref{relicobserve}, corresponding to the black dashed line in the graph. The red dots indicate the points where the calculated value equals the observed value~\eqref{relicobserve} and BPs are included.

On the other hand, the right panel of Fig.~\ref{fig:DM} shows the $m_\chi$ dependence of the spin-independent scattering cross section between DM and nucleus. For $\Omega_\chi<\Omega_{\mathrm{DM}}$, we should scale $\sigma_{\mathrm{SI}}$ as
\begin{align}
\tilde{\sigma}_{\mathrm{SI}}=\left(\frac{\Omega_\chi}{\Omega_{\mathrm{DM}}}\right) \sigma_{\mathrm{SI}}.
\end{align}
Here, the scaled cross section is shown for each BP. The dotted line is the bound of the LZ experiment, and the region above this is excluded. The red dots correspond to those of the left panel. It can be seen that many DM mass regions are still allowed, and the DM $\chi$ may carry all of the existing DM and explain the current experimental results.


\section{Summary and discussion}\label{sec:sum}
In this paper, we have investigated the compatibility of the degenerate scalar scenario and strong first-order EWPT in the 2HDMS, which contains two Higgs doublets and one singlet scalar. In this model, the imaginary part of $S$ behaves as a DM $\chi$, and CP-even scalars mix to form three Higgs particles $H_i~(i=1,2,3)$.
The scattering of DM and quarks is mediated by $H_i$, and we have found that this scattering is suppressed in the region where Higgs masses are degenerate, as shown in previous studies on pNG DM~\cite{Abe:2021nih}.
The suppression of SU(2)$_L$ doublet-singlet mixing terms $\delta_1$ and $\delta_2$, which are related to the orthogonality of the mixing matrix that transforms the gauge eigenstates to the mass eigenstates of the CP-even scalars, is important for the degenerate scalar scenario.

Strong first-order EWPT, an essential element of EWBG, can be derived from two sources in the effective potential: the thermal loop and the tree-level structure. Since the tree-level driven EWPT is induced by the presence of the singlet field, i.e., large $\delta_1,~\delta_2$ are important, it was pointed out here that this is inconsistent with the degenerate scalar scenario (the discussion in the CxSM is given in ref.~\cite{Cho:2021itv}). Therefore, as is common with the 2HDM, it is best to leave strong first-order EWPT to the loop effect.

We have set the benchmark points based on the qualitative discussion of EWPT and confirmed that strong first-order EWPT actually occurs. Since we focus on the degenerate scalar scenario, the DM-quark spin-independent cross section is suppressed even in regions that satisfy the DM observed relic density, and we were able to explain the constraint from the LZ experiment in many regions. Therefore, it was found that current observational data and experimental results can be explained by the DM $\chi$ in this model and that it is also consistent with the condition given by strong first-order EWPT.

\begin{table}[t]
\center
\begin{tabular}{|c|c|c|c|c|}
\hline &  $\lambda_{111}/\lambda_{hhh}^{\mathrm{SM}}
$ & $\lambda_{112}/\lambda_{hhh}^{\mathrm{SM}}
$ & $\lambda_{113}/\lambda_{hhh}^{\mathrm{SM}}
$ & $\Delta\lambda_{11}^\mathrm{total}$  \\
\hline BP1 &  $0.799$ & $-0.00872$ & $0.571$ & 0.362 \\
\hline BP2 & $0.780$ & $0.0109$ & $0.544$ & 0.335\\
\hline BP3 & $0.701$ & $0.0921$ & $0.431$ & 0.224\\
\hline
\end{tabular}
\caption{The deviation from the SM values for the Higgs trilinear couplings $\lambda_{111}$, $\lambda_{112}$, $\lambda_{113}$ and their sum defined as Eq.~\eqref{trlnr}.
}
\label{tab:selfcoupling}
\end{table}
 
Finally, we briefly discuss the possibility of verifying the model in future collider experiments. 
A distinctive feature of the thermal loop-derived EWPT is a shift of the Higgs trilinear coupling
\begin{align}
 \mathcal{L} &= \lambda_{ijk} H_i H_j H_k 
~~~(i,j,k=1,2,3), 
\end{align}
from the SM prediction~\cite{Kanemura:2004ch,Grojean:2004xa}. For example, 
we focus on the pair production of $H_1$ through off-shell CP-even scalars $H_i$, i.e., $H_i \to H_1 H_1$. 
We compare each trilinear coupling $\lambda_{11i}~(i=1,2,3)$ in three benchmark points (BP1, BP2, BP3) with the SM. Furthermore,  
since amplitudes mediated by off-shell scalars $H_i$ interfere with each other, we also compare the sum of trilinear couplings $\lambda_{111}, \lambda_{112}$ and $\lambda_{113}$ to the SM prediction. 
The results are shown in Table.~\ref{tab:selfcoupling}, where $\Delta \lambda_{11}^{\mathrm{total}}$ is defined as 
\begin{align}
\Delta\lambda_{11}^\mathrm{total}
=\frac{(\lambda_{111}+\lambda_{112}+\lambda_{113})-\lambda^\mathrm{SM}_{hhh}}{\lambda^\mathrm{SM}_{hhh}},
\label{trlnr}
\end{align}
The Higgs trilinear coupling of this model is expected to differ from that of SM by about 30\%, and it may be confirmed at the High-Energy Large Hadron Collider (HE-LHC) and/or the International Linear Collider (ILC), whose accuracy is estimated by ref.~\cite{Shiltsev:2019szl}. We leave the detailed analysis and the model verification in other experiments, such as the search for the degenerate Higgs at the lepton collider and observation of gravitational waves from EWPT, to future research.


\begin{acknowledgments}
We are grateful to Ayana Okuma and Chiaki Nose for their valuable discussions. The work of GCC is supported by JSPS KAKENHI Grant No. 22K03616. The part of the work of CI is supported by the National Natural Science Foundation of China (NNSFC) Grant No. 12205387, No.12475111.

\end{acknowledgments}

\appendix

\section{Input and output parameters}\label{app:para}
The following shows the relationship between the input and output parameters. 
However, $m_3^2$ and $a_1$ remain as input parameters. 
The index $i$ runs from 1 to 3, and the sum symbol $\qty(\sum_i)$ is omitted.
\begin{align}
m_1^2 &= \frac{m_3^2 v_2}{v_1} - \frac{\lambda_1 v_1^2}{2} - \frac{\lambda_{345} v_2^2}{2} -\frac{\delta_1 v_S^2}{4}, \\
m_2^2 &= \frac{m_3^2 v_1}{v_2} - \frac{\lambda_2 v_2^2}{2} - \frac{\lambda_{345} v_1^2}{2} -\frac{\delta_2 v_S^2}{4}, \\
\lambda_1 &= \frac{1}{v_1^2} \left(\sum_{i=1}^3 O_{1i}^2 m_{H_i}^2 - \frac{m_3^2 v_2}{v_1} \right),\\
\lambda_2 &= \frac{1}{v_2^2} \left(\sum_{i=1}^3 O_{2i}^2 m_{H_i}^2- \frac{m_3^2 v_1}{v_2} \right),\\
\lambda_3 &= \frac{1}{v_1 v_2} \left(\sum_{i=1}^3 O_{1i} O_{2i} m_{H_i}^2  + m_3^2 \right) - \lambda_{4} - \lambda_{5}, \\
\lambda_{4} &= \frac{2}{v^2} \left( \frac{m_3^2}{\sin\beta \cos\beta} - m_{H^\pm}^2 \right) - \lambda_{5}, \\
\lambda_{5} &= \frac{1}{v^2}\left(\frac{m_3^2}{\sin\beta \cos\beta} - m_A^2\right),\\
\delta_1 &= \frac{2}{v_1 v_S} \sum_{i=1}^3 O_{1i} O_{3i} m_{H_i}^2 , \\
\delta_2 &= \frac{2}{v_2 v_S} \sum_{i=1}^3 O_{2i} O_{3i} m_{H_i}^2 , \\
d_2 &= \frac{2}{v_S^2} \left( \frac{\sqrt{2} a_1}{v_S} + \sum_{i=1}^3 O_{3i}^2 m_{H_i}^2 \right), \\
b_2 &= \frac{-4 \sqrt{2} a_1 - 2 b_1 v_S - \delta_1 v_1^2 v_S - \delta_2 v_2^2 v_S - d_2 v_S^3}{2 v_S}, \\
b_1 &= -m_{\chi}^2 - \frac{\sqrt{2} a_1}{v_S}.
\end{align}

\section{Trilinear and quartic couplings in the pNG-DM model}\label{app:bitri}
We collect trilinear and bilinear couplings of the potential \eqref{pot} in the gauge eigenstate, which appear in Sec.~\ref{sec:dege}.
\begin{align}
C_{\chi\chi h_1}&=\frac{\delta_1}{4}v_1,\quad C_{\chi\chi h_2}=\frac{\delta_2}{4}v_2,\quad C_{\chi\chi s}=\frac{d_2}{4}v_S,\nonumber \\
C_{h_1 s}&=\frac{\delta_2}{4}v_1 v_S,\quad C_{h_2 s}=\frac{\delta_2}{2}v_2 v_S,\quad C_{s s}=\frac{b_1}{2}+\frac{b_2}{2}+\frac{\delta_1}{8}v_1^2+\frac{\delta_1}{8}v_2^2+\frac{3 d_2}{8}v_S^2.
\end{align}

\bibliography{biblist}

\begin{thebibliography}{49}%
\makeatletter
\providecommand \@ifxundefined [1]{%
 \@ifx{#1\undefined}
}%
\providecommand \@ifnum [1]{%
 \ifnum #1\expandafter \@firstoftwo
 \else \expandafter \@secondoftwo
 \fi
}%
\providecommand \@ifx [1]{%
 \ifx #1\expandafter \@firstoftwo
 \else \expandafter \@secondoftwo
 \fi
}%
\providecommand \natexlab [1]{#1}%
\providecommand \enquote  [1]{``#1''}%
\providecommand \bibnamefont  [1]{#1}%
\providecommand \bibfnamefont [1]{#1}%
\providecommand \citenamefont [1]{#1}%
\providecommand \href@noop [0]{\@secondoftwo}%
\providecommand \href [0]{\begingroup \@sanitize@url \@href}%
\providecommand \@href[1]{\@@startlink{#1}\@@href}%
\providecommand \@@href[1]{\endgroup#1\@@endlink}%
\providecommand \@sanitize@url [0]{\catcode `\\12\catcode `\$12\catcode
  `\&12\catcode `\#12\catcode `\^12\catcode `\_12\catcode `\%12\relax}%
\providecommand \@@startlink[1]{}%
\providecommand \@@endlink[0]{}%
\providecommand \url  [0]{\begingroup\@sanitize@url \@url }%
\providecommand \@url [1]{\endgroup\@href {#1}{\urlprefix }}%
\providecommand \urlprefix  [0]{URL }%
\providecommand \Eprint [0]{\href }%
\providecommand \doibase [0]{http://dx.doi.org/}%
\providecommand \selectlanguage [0]{\@gobble}%
\providecommand \bibinfo  [0]{\@secondoftwo}%
\providecommand \bibfield  [0]{\@secondoftwo}%
\providecommand \translation [1]{[#1]}%
\providecommand \BibitemOpen [0]{}%
\providecommand \bibitemStop [0]{}%
\providecommand \bibitemNoStop [0]{.\EOS\space}%
\providecommand \EOS [0]{\spacefactor3000\relax}%
\providecommand \BibitemShut  [1]{\csname bibitem#1\endcsname}%
\let\auto@bib@innerbib\@empty
\bibitem [{\citenamefont {Aalbers}\ \emph {et~al.}(2024)\citenamefont {Aalbers}
  \emph {et~al.}}]{LZ:2024zvo}%
  \BibitemOpen
  \bibfield  {author} {\bibinfo {author} {\bibfnamefont {J.}~\bibnamefont
  {Aalbers}} \emph {et~al.} (\bibinfo {collaboration} {LZ}),\ }\href@noop {} {\
   (\bibinfo {year} {2024})},\ \Eprint {http://arxiv.org/abs/2410.17036}
  {arXiv:2410.17036 [hep-ex]} \BibitemShut {NoStop}%
\bibitem [{\citenamefont {Barger}\ \emph {et~al.}(2009)\citenamefont {Barger},
  \citenamefont {Langacker}, \citenamefont {McCaskey}, \citenamefont
  {Ramsey-Musolf},\ and\ \citenamefont {Shaughnessy}}]{Barger:2008jx}%
  \BibitemOpen
  \bibfield  {author} {\bibinfo {author} {\bibfnamefont {V.}~\bibnamefont
  {Barger}}, \bibinfo {author} {\bibfnamefont {P.}~\bibnamefont {Langacker}},
  \bibinfo {author} {\bibfnamefont {M.}~\bibnamefont {McCaskey}}, \bibinfo
  {author} {\bibfnamefont {M.}~\bibnamefont {Ramsey-Musolf}}, \ and\ \bibinfo
  {author} {\bibfnamefont {G.}~\bibnamefont {Shaughnessy}},\ }\href {\doibase
  10.1103/PhysRevD.79.015018} {\bibfield  {journal} {\bibinfo  {journal} {Phys.
  Rev. D}\ }\textbf {\bibinfo {volume} {79}},\ \bibinfo {pages} {015018}
  (\bibinfo {year} {2009})},\ \Eprint {http://arxiv.org/abs/0811.0393}
  {arXiv:0811.0393 [hep-ph]} \BibitemShut {NoStop}%
\bibitem [{\citenamefont {Gross}\ \emph {et~al.}(2017)\citenamefont {Gross},
  \citenamefont {Lebedev},\ and\ \citenamefont {Toma}}]{Gross:2017dan}%
  \BibitemOpen
  \bibfield  {author} {\bibinfo {author} {\bibfnamefont {C.}~\bibnamefont
  {Gross}}, \bibinfo {author} {\bibfnamefont {O.}~\bibnamefont {Lebedev}}, \
  and\ \bibinfo {author} {\bibfnamefont {T.}~\bibnamefont {Toma}},\ }\href
  {\doibase 10.1103/PhysRevLett.119.191801} {\bibfield  {journal} {\bibinfo
  {journal} {Phys. Rev. Lett.}\ }\textbf {\bibinfo {volume} {119}},\ \bibinfo
  {pages} {191801} (\bibinfo {year} {2017})},\ \Eprint
  {http://arxiv.org/abs/1708.02253} {arXiv:1708.02253 [hep-ph]} \BibitemShut
  {NoStop}%
\bibitem [{\citenamefont {Abe}\ \emph {et~al.}(2021)\citenamefont {Abe},
  \citenamefont {Cho},\ and\ \citenamefont {Mawatari}}]{Abe:2021nih}%
  \BibitemOpen
  \bibfield  {author} {\bibinfo {author} {\bibfnamefont {S.}~\bibnamefont
  {Abe}}, \bibinfo {author} {\bibfnamefont {G.-C.}\ \bibnamefont {Cho}}, \ and\
  \bibinfo {author} {\bibfnamefont {K.}~\bibnamefont {Mawatari}},\ }\href
  {\doibase 10.1103/PhysRevD.104.035023} {\bibfield  {journal} {\bibinfo
  {journal} {Phys. Rev. D}\ }\textbf {\bibinfo {volume} {104}},\ \bibinfo
  {pages} {035023} (\bibinfo {year} {2021})},\ \Eprint
  {http://arxiv.org/abs/2101.04887} {arXiv:2101.04887 [hep-ph]} \BibitemShut
  {NoStop}%
\bibitem [{\citenamefont {Kuzmin}\ \emph {et~al.}(1985)\citenamefont {Kuzmin},
  \citenamefont {Rubakov},\ and\ \citenamefont {Shaposhnikov}}]{Kuzmin:1985mm}%
  \BibitemOpen
  \bibfield  {author} {\bibinfo {author} {\bibfnamefont {V.~A.}\ \bibnamefont
  {Kuzmin}}, \bibinfo {author} {\bibfnamefont {V.~A.}\ \bibnamefont {Rubakov}},
  \ and\ \bibinfo {author} {\bibfnamefont {M.~E.}\ \bibnamefont
  {Shaposhnikov}},\ }\href {\doibase 10.1016/0370-2693(85)91028-7} {\bibfield
  {journal} {\bibinfo  {journal} {Phys. Lett.}\ }\textbf {\bibinfo {volume}
  {155B}},\ \bibinfo {pages} {36} (\bibinfo {year} {1985})}\BibitemShut
  {NoStop}%
\bibitem [{\citenamefont {Rubakov}\ and\ \citenamefont
  {Shaposhnikov}(1996)}]{Rubakov:1996vz}%
  \BibitemOpen
  \bibfield  {author} {\bibinfo {author} {\bibfnamefont {V.~A.}\ \bibnamefont
  {Rubakov}}\ and\ \bibinfo {author} {\bibfnamefont {M.~E.}\ \bibnamefont
  {Shaposhnikov}},\ }\href {\doibase 10.1070/PU1996v039n05ABEH000145}
  {\bibfield  {journal} {\bibinfo  {journal} {Usp. Fiz. Nauk}\ }\textbf
  {\bibinfo {volume} {166}},\ \bibinfo {pages} {493} (\bibinfo {year}
  {1996})},\ \bibinfo {note} {[Phys. Usp.39,461(1996)]},\ \Eprint
  {http://arxiv.org/abs/hep-ph/9603208} {arXiv:hep-ph/9603208 [hep-ph]}
  \BibitemShut {NoStop}%
\bibitem [{\citenamefont {Funakubo}(1996)}]{Funakubo:1996dw}%
  \BibitemOpen
  \bibfield  {author} {\bibinfo {author} {\bibfnamefont {K.}~\bibnamefont
  {Funakubo}},\ }\href {\doibase 10.1143/PTP.96.475} {\bibfield  {journal}
  {\bibinfo  {journal} {Prog. Theor. Phys.}\ }\textbf {\bibinfo {volume}
  {96}},\ \bibinfo {pages} {475} (\bibinfo {year} {1996})},\ \Eprint
  {http://arxiv.org/abs/hep-ph/9608358} {arXiv:hep-ph/9608358 [hep-ph]}
  \BibitemShut {NoStop}%
\bibitem [{\citenamefont {Riotto}(1998)}]{Riotto:1998bt}%
  \BibitemOpen
  \bibfield  {author} {\bibinfo {author} {\bibfnamefont {A.}~\bibnamefont
  {Riotto}},\ }in\ \href@noop {} {\emph {\bibinfo {booktitle} {{Proceedings,
  Summer School in High-energy physics and cosmology: Trieste, Italy, June
  29-July 17, 1998}}}}\ (\bibinfo {year} {1998})\ pp.\ \bibinfo {pages}
  {326--436},\ \Eprint {http://arxiv.org/abs/hep-ph/9807454}
  {arXiv:hep-ph/9807454 [hep-ph]} \BibitemShut {NoStop}%
\bibitem [{\citenamefont {Trodden}(1999)}]{Trodden:1998ym}%
  \BibitemOpen
  \bibfield  {author} {\bibinfo {author} {\bibfnamefont {M.}~\bibnamefont
  {Trodden}},\ }\href {\doibase 10.1103/RevModPhys.71.1463} {\bibfield
  {journal} {\bibinfo  {journal} {Rev. Mod. Phys.}\ }\textbf {\bibinfo {volume}
  {71}},\ \bibinfo {pages} {1463} (\bibinfo {year} {1999})},\ \Eprint
  {http://arxiv.org/abs/hep-ph/9803479} {arXiv:hep-ph/9803479 [hep-ph]}
  \BibitemShut {NoStop}%
\bibitem [{\citenamefont {Bernreuther}(2002)}]{Bernreuther:2002uj}%
  \BibitemOpen
  \bibfield  {author} {\bibinfo {author} {\bibfnamefont {W.}~\bibnamefont
  {Bernreuther}},\ }\bibfield  {booktitle} {\emph {\bibinfo {booktitle}
  {{Workshop of the Graduate College of Elementary Particle Physics Berlin,
  Germany, April 2-5, 2001}}},\ }\href@noop {} {\bibfield  {journal} {\bibinfo
  {journal} {Lect. Notes Phys.}\ }\textbf {\bibinfo {volume} {591}},\ \bibinfo
  {pages} {237} (\bibinfo {year} {2002})},\ \bibinfo {note} {[,237(2002)]},\
  \Eprint {http://arxiv.org/abs/hep-ph/0205279} {arXiv:hep-ph/0205279 [hep-ph]}
  \BibitemShut {NoStop}%
\bibitem [{\citenamefont {Cline}(2006)}]{Cline:2006ts}%
  \BibitemOpen
  \bibfield  {author} {\bibinfo {author} {\bibfnamefont {J.~M.}\ \bibnamefont
  {Cline}},\ }in\ \href@noop {} {\emph {\bibinfo {booktitle} {{Les Houches
  Summer School - Session 86: Particle Physics and Cosmology: The Fabric of
  Spacetime Les Houches, France, July 31-August 25, 2006}}}}\ (\bibinfo {year}
  {2006})\ \Eprint {http://arxiv.org/abs/hep-ph/0609145} {arXiv:hep-ph/0609145
  [hep-ph]} \BibitemShut {NoStop}%
\bibitem [{\citenamefont {Morrissey}\ and\ \citenamefont
  {Ramsey-Musolf}(2012)}]{Morrissey:2012db}%
  \BibitemOpen
  \bibfield  {author} {\bibinfo {author} {\bibfnamefont {D.~E.}\ \bibnamefont
  {Morrissey}}\ and\ \bibinfo {author} {\bibfnamefont {M.~J.}\ \bibnamefont
  {Ramsey-Musolf}},\ }\href {\doibase 10.1088/1367-2630/14/12/125003}
  {\bibfield  {journal} {\bibinfo  {journal} {New J. Phys.}\ }\textbf {\bibinfo
  {volume} {14}},\ \bibinfo {pages} {125003} (\bibinfo {year} {2012})},\
  \Eprint {http://arxiv.org/abs/1206.2942} {arXiv:1206.2942 [hep-ph]}
  \BibitemShut {NoStop}%
\bibitem [{\citenamefont {Konstandin}(2013)}]{Konstandin:2013caa}%
  \BibitemOpen
  \bibfield  {author} {\bibinfo {author} {\bibfnamefont {T.}~\bibnamefont
  {Konstandin}},\ }\href {\doibase 10.3367/UFNe.0183.201308a.0785} {\bibfield
  {journal} {\bibinfo  {journal} {Phys. Usp.}\ }\textbf {\bibinfo {volume}
  {56}},\ \bibinfo {pages} {747} (\bibinfo {year} {2013})},\ \bibinfo {note}
  {[Usp. Fiz. Nauk183,785(2013)]},\ \Eprint {http://arxiv.org/abs/1302.6713}
  {arXiv:1302.6713 [hep-ph]} \BibitemShut {NoStop}%
\bibitem [{\citenamefont {Senaha}(2020)}]{Senaha:2020mop}%
  \BibitemOpen
  \bibfield  {author} {\bibinfo {author} {\bibfnamefont {E.}~\bibnamefont
  {Senaha}},\ }\href {\doibase 10.3390/sym12050733} {\bibfield  {journal}
  {\bibinfo  {journal} {Symmetry}\ }\textbf {\bibinfo {volume} {12}},\ \bibinfo
  {pages} {733} (\bibinfo {year} {2020})}\BibitemShut {NoStop}%
\bibitem [{\citenamefont {Kajantie}\ \emph {et~al.}(1996)\citenamefont
  {Kajantie}, \citenamefont {Laine}, \citenamefont {Rummukainen},\ and\
  \citenamefont {Shaposhnikov}}]{Kajantie:1996mn}%
  \BibitemOpen
  \bibfield  {author} {\bibinfo {author} {\bibfnamefont {K.}~\bibnamefont
  {Kajantie}}, \bibinfo {author} {\bibfnamefont {M.}~\bibnamefont {Laine}},
  \bibinfo {author} {\bibfnamefont {K.}~\bibnamefont {Rummukainen}}, \ and\
  \bibinfo {author} {\bibfnamefont {M.~E.}\ \bibnamefont {Shaposhnikov}},\
  }\href {\doibase 10.1103/PhysRevLett.77.2887} {\bibfield  {journal} {\bibinfo
   {journal} {Phys. Rev. Lett.}\ }\textbf {\bibinfo {volume} {77}},\ \bibinfo
  {pages} {2887} (\bibinfo {year} {1996})},\ \Eprint
  {http://arxiv.org/abs/hep-ph/9605288} {arXiv:hep-ph/9605288} \BibitemShut
  {NoStop}%
\bibitem [{\citenamefont {Rummukainen}\ \emph {et~al.}(1998)\citenamefont
  {Rummukainen}, \citenamefont {Tsypin}, \citenamefont {Kajantie},
  \citenamefont {Laine},\ and\ \citenamefont
  {Shaposhnikov}}]{Rummukainen:1998as}%
  \BibitemOpen
  \bibfield  {author} {\bibinfo {author} {\bibfnamefont {K.}~\bibnamefont
  {Rummukainen}}, \bibinfo {author} {\bibfnamefont {M.}~\bibnamefont {Tsypin}},
  \bibinfo {author} {\bibfnamefont {K.}~\bibnamefont {Kajantie}}, \bibinfo
  {author} {\bibfnamefont {M.}~\bibnamefont {Laine}}, \ and\ \bibinfo {author}
  {\bibfnamefont {M.~E.}\ \bibnamefont {Shaposhnikov}},\ }\href {\doibase
  10.1016/S0550-3213(98)00494-5} {\bibfield  {journal} {\bibinfo  {journal}
  {Nucl. Phys. B}\ }\textbf {\bibinfo {volume} {532}},\ \bibinfo {pages} {283}
  (\bibinfo {year} {1998})},\ \Eprint {http://arxiv.org/abs/hep-lat/9805013}
  {arXiv:hep-lat/9805013} \BibitemShut {NoStop}%
\bibitem [{\citenamefont {Csikor}\ \emph {et~al.}(1999)\citenamefont {Csikor},
  \citenamefont {Fodor},\ and\ \citenamefont {Heitger}}]{Csikor:1998eu}%
  \BibitemOpen
  \bibfield  {author} {\bibinfo {author} {\bibfnamefont {F.}~\bibnamefont
  {Csikor}}, \bibinfo {author} {\bibfnamefont {Z.}~\bibnamefont {Fodor}}, \
  and\ \bibinfo {author} {\bibfnamefont {J.}~\bibnamefont {Heitger}},\ }\href
  {\doibase 10.1103/PhysRevLett.82.21} {\bibfield  {journal} {\bibinfo
  {journal} {Phys. Rev. Lett.}\ }\textbf {\bibinfo {volume} {82}},\ \bibinfo
  {pages} {21} (\bibinfo {year} {1999})},\ \Eprint
  {http://arxiv.org/abs/hep-ph/9809291} {arXiv:hep-ph/9809291} \BibitemShut
  {NoStop}%
\bibitem [{\citenamefont {Aoki}\ \emph {et~al.}(1999)\citenamefont {Aoki},
  \citenamefont {Csikor}, \citenamefont {Fodor},\ and\ \citenamefont
  {Ukawa}}]{Aoki:1999fi}%
  \BibitemOpen
  \bibfield  {author} {\bibinfo {author} {\bibfnamefont {Y.}~\bibnamefont
  {Aoki}}, \bibinfo {author} {\bibfnamefont {F.}~\bibnamefont {Csikor}},
  \bibinfo {author} {\bibfnamefont {Z.}~\bibnamefont {Fodor}}, \ and\ \bibinfo
  {author} {\bibfnamefont {A.}~\bibnamefont {Ukawa}},\ }\href {\doibase
  10.1103/PhysRevD.60.013001} {\bibfield  {journal} {\bibinfo  {journal} {Phys.
  Rev. D}\ }\textbf {\bibinfo {volume} {60}},\ \bibinfo {pages} {013001}
  (\bibinfo {year} {1999})},\ \Eprint {http://arxiv.org/abs/hep-lat/9901021}
  {arXiv:hep-lat/9901021} \BibitemShut {NoStop}%
\bibitem [{\citenamefont {Cho}\ \emph {et~al.}(2021)\citenamefont {Cho},
  \citenamefont {Idegawa},\ and\ \citenamefont {Senaha}}]{Cho:2021itv}%
  \BibitemOpen
  \bibfield  {author} {\bibinfo {author} {\bibfnamefont {G.-C.}\ \bibnamefont
  {Cho}}, \bibinfo {author} {\bibfnamefont {C.}~\bibnamefont {Idegawa}}, \ and\
  \bibinfo {author} {\bibfnamefont {E.}~\bibnamefont {Senaha}},\ }\href
  {\doibase 10.1016/j.physletb.2021.136787} {\bibfield  {journal} {\bibinfo
  {journal} {Phys. Lett. B}\ }\textbf {\bibinfo {volume} {823}},\ \bibinfo
  {pages} {136787} (\bibinfo {year} {2021})},\ \Eprint
  {http://arxiv.org/abs/2105.11830} {arXiv:2105.11830 [hep-ph]} \BibitemShut
  {NoStop}%
\bibitem [{\citenamefont {Aghanim}\ \emph {et~al.}(2020)\citenamefont {Aghanim}
  \emph {et~al.}}]{Planck:2018vyg}%
  \BibitemOpen
  \bibfield  {author} {\bibinfo {author} {\bibfnamefont {N.}~\bibnamefont
  {Aghanim}} \emph {et~al.} (\bibinfo {collaboration} {Planck}),\ }\href
  {\doibase 10.1051/0004-6361/201833910} {\bibfield  {journal} {\bibinfo
  {journal} {Astron. Astrophys.}\ }\textbf {\bibinfo {volume} {641}},\ \bibinfo
  {pages} {A6} (\bibinfo {year} {2020})},\ \bibinfo {note} {[Erratum:
  Astron.Astrophys. 652, C4 (2021)]},\ \Eprint
  {http://arxiv.org/abs/1807.06209} {arXiv:1807.06209 [astro-ph.CO]}
  \BibitemShut {NoStop}%
\bibitem [{\citenamefont {Jiang}\ \emph {et~al.}(2019)\citenamefont {Jiang},
  \citenamefont {Cai}, \citenamefont {Yu}, \citenamefont {Zeng},\ and\
  \citenamefont {Zhang}}]{Jiang:2019soj}%
  \BibitemOpen
  \bibfield  {author} {\bibinfo {author} {\bibfnamefont {X.-M.}\ \bibnamefont
  {Jiang}}, \bibinfo {author} {\bibfnamefont {C.}~\bibnamefont {Cai}}, \bibinfo
  {author} {\bibfnamefont {Z.-H.}\ \bibnamefont {Yu}}, \bibinfo {author}
  {\bibfnamefont {Y.-P.}\ \bibnamefont {Zeng}}, \ and\ \bibinfo {author}
  {\bibfnamefont {H.-H.}\ \bibnamefont {Zhang}},\ }\href {\doibase
  10.1103/PhysRevD.100.075011} {\bibfield  {journal} {\bibinfo  {journal}
  {Phys. Rev. D}\ }\textbf {\bibinfo {volume} {100}},\ \bibinfo {pages}
  {075011} (\bibinfo {year} {2019})},\ \Eprint
  {http://arxiv.org/abs/1907.09684} {arXiv:1907.09684 [hep-ph]} \BibitemShut
  {NoStop}%
\bibitem [{\citenamefont {Zhang}\ \emph {et~al.}(2021)\citenamefont {Zhang},
  \citenamefont {Cai}, \citenamefont {Jiang}, \citenamefont {Tang},
  \citenamefont {Yu},\ and\ \citenamefont {Zhang}}]{Zhang:2021alu}%
  \BibitemOpen
  \bibfield  {author} {\bibinfo {author} {\bibfnamefont {Z.}~\bibnamefont
  {Zhang}}, \bibinfo {author} {\bibfnamefont {C.}~\bibnamefont {Cai}}, \bibinfo
  {author} {\bibfnamefont {X.-M.}\ \bibnamefont {Jiang}}, \bibinfo {author}
  {\bibfnamefont {Y.-L.}\ \bibnamefont {Tang}}, \bibinfo {author}
  {\bibfnamefont {Z.-H.}\ \bibnamefont {Yu}}, \ and\ \bibinfo {author}
  {\bibfnamefont {H.-H.}\ \bibnamefont {Zhang}},\ }\href {\doibase
  10.1007/JHEP05(2021)160} {\bibfield  {journal} {\bibinfo  {journal} {JHEP}\
  }\textbf {\bibinfo {volume} {05}},\ \bibinfo {pages} {160} (\bibinfo {year}
  {2021})},\ \Eprint {http://arxiv.org/abs/2102.01588} {arXiv:2102.01588
  [hep-ph]} \BibitemShut {NoStop}%
\bibitem [{\citenamefont {Biek\"otter}\ and\ \citenamefont
  {Olea-Romacho}(2021)}]{Biekotter:2021ovi}%
  \BibitemOpen
  \bibfield  {author} {\bibinfo {author} {\bibfnamefont {T.}~\bibnamefont
  {Biek\"otter}}\ and\ \bibinfo {author} {\bibfnamefont {M.~O.}\ \bibnamefont
  {Olea-Romacho}},\ }\href {\doibase 10.1007/JHEP10(2021)215} {\bibfield
  {journal} {\bibinfo  {journal} {JHEP}\ }\textbf {\bibinfo {volume} {10}},\
  \bibinfo {pages} {215} (\bibinfo {year} {2021})},\ \Eprint
  {http://arxiv.org/abs/2108.10864} {arXiv:2108.10864 [hep-ph]} \BibitemShut
  {NoStop}%
\bibitem [{\citenamefont {Biek\"otter}\ \emph {et~al.}(2022)\citenamefont
  {Biek\"otter}, \citenamefont {Gabriel}, \citenamefont {Olea-Romacho},\ and\
  \citenamefont {Santos}}]{Biekotter:2022bxp}%
  \BibitemOpen
  \bibfield  {author} {\bibinfo {author} {\bibfnamefont {T.}~\bibnamefont
  {Biek\"otter}}, \bibinfo {author} {\bibfnamefont {P.}~\bibnamefont
  {Gabriel}}, \bibinfo {author} {\bibfnamefont {M.~O.}\ \bibnamefont
  {Olea-Romacho}}, \ and\ \bibinfo {author} {\bibfnamefont {R.}~\bibnamefont
  {Santos}},\ }\href {\doibase 10.1007/JHEP10(2022)126} {\bibfield  {journal}
  {\bibinfo  {journal} {JHEP}\ }\textbf {\bibinfo {volume} {10}},\ \bibinfo
  {pages} {126} (\bibinfo {year} {2022})},\ \Eprint
  {http://arxiv.org/abs/2207.04973} {arXiv:2207.04973 [hep-ph]} \BibitemShut
  {NoStop}%
\bibitem [{\citenamefont {Cho}\ \emph {et~al.}(2024)\citenamefont {Cho},
  \citenamefont {Idegawa},\ and\ \citenamefont {Inumiya}}]{Cho:2023oad}%
  \BibitemOpen
  \bibfield  {author} {\bibinfo {author} {\bibfnamefont {G.-C.}\ \bibnamefont
  {Cho}}, \bibinfo {author} {\bibfnamefont {C.}~\bibnamefont {Idegawa}}, \ and\
  \bibinfo {author} {\bibfnamefont {R.}~\bibnamefont {Inumiya}},\ }\href
  {\doibase 10.1016/j.nuclphysb.2024.116688} {\bibfield  {journal} {\bibinfo
  {journal} {Nucl. Phys. B}\ }\textbf {\bibinfo {volume} {1007}},\ \bibinfo
  {pages} {116688} (\bibinfo {year} {2024})},\ \Eprint
  {http://arxiv.org/abs/2312.05776} {arXiv:2312.05776 [hep-ph]} \BibitemShut
  {NoStop}%
\bibitem [{\citenamefont {Branco}\ \emph {et~al.}(2012)\citenamefont {Branco},
  \citenamefont {Ferreira}, \citenamefont {Lavoura}, \citenamefont {Rebelo},
  \citenamefont {Sher},\ and\ \citenamefont {Silva}}]{Branco:2011iw}%
  \BibitemOpen
  \bibfield  {author} {\bibinfo {author} {\bibfnamefont {G.~C.}\ \bibnamefont
  {Branco}}, \bibinfo {author} {\bibfnamefont {P.~M.}\ \bibnamefont
  {Ferreira}}, \bibinfo {author} {\bibfnamefont {L.}~\bibnamefont {Lavoura}},
  \bibinfo {author} {\bibfnamefont {M.~N.}\ \bibnamefont {Rebelo}}, \bibinfo
  {author} {\bibfnamefont {M.}~\bibnamefont {Sher}}, \ and\ \bibinfo {author}
  {\bibfnamefont {J.~P.}\ \bibnamefont {Silva}},\ }\href {\doibase
  10.1016/j.physrep.2012.02.002} {\bibfield  {journal} {\bibinfo  {journal}
  {Phys. Rept.}\ }\textbf {\bibinfo {volume} {516}},\ \bibinfo {pages} {1}
  (\bibinfo {year} {2012})},\ \Eprint {http://arxiv.org/abs/1106.0034}
  {arXiv:1106.0034 [hep-ph]} \BibitemShut {NoStop}%
\bibitem [{\citenamefont {Cho}\ and\ \citenamefont
  {Idegawa}(2023)}]{Cho:2023hek}%
  \BibitemOpen
  \bibfield  {author} {\bibinfo {author} {\bibfnamefont {G.-C.}\ \bibnamefont
  {Cho}}\ and\ \bibinfo {author} {\bibfnamefont {C.}~\bibnamefont {Idegawa}},\
  }\href {\doibase 10.1016/j.nuclphysb.2023.116320} {\bibfield  {journal}
  {\bibinfo  {journal} {Nucl. Phys. B}\ }\textbf {\bibinfo {volume} {994}},\
  \bibinfo {pages} {116320} (\bibinfo {year} {2023})},\ \Eprint
  {http://arxiv.org/abs/2304.10096} {arXiv:2304.10096 [hep-ph]} \BibitemShut
  {NoStop}%
\bibitem [{\citenamefont {Arnold}\ and\ \citenamefont
  {McLerran}(1987)}]{Arnold:1987mh}%
  \BibitemOpen
  \bibfield  {author} {\bibinfo {author} {\bibfnamefont {P.~B.}\ \bibnamefont
  {Arnold}}\ and\ \bibinfo {author} {\bibfnamefont {L.~D.}\ \bibnamefont
  {McLerran}},\ }\href {\doibase 10.1103/PhysRevD.36.581} {\bibfield  {journal}
  {\bibinfo  {journal} {Phys. Rev. D}\ }\textbf {\bibinfo {volume} {36}},\
  \bibinfo {pages} {581} (\bibinfo {year} {1987})}\BibitemShut {NoStop}%
\bibitem [{\citenamefont {Bochkarev}\ and\ \citenamefont
  {Shaposhnikov}(1987)}]{Bochkarev:1987wf}%
  \BibitemOpen
  \bibfield  {author} {\bibinfo {author} {\bibfnamefont {A.~I.}\ \bibnamefont
  {Bochkarev}}\ and\ \bibinfo {author} {\bibfnamefont {M.~E.}\ \bibnamefont
  {Shaposhnikov}},\ }\href {\doibase 10.1142/S0217732387000537} {\bibfield
  {journal} {\bibinfo  {journal} {Mod. Phys. Lett. A}\ }\textbf {\bibinfo
  {volume} {2}},\ \bibinfo {pages} {417} (\bibinfo {year} {1987})}\BibitemShut
  {NoStop}%
\bibitem [{\citenamefont {Funakubo}\ and\ \citenamefont
  {Senaha}(2009)}]{Funakubo:2009eg}%
  \BibitemOpen
  \bibfield  {author} {\bibinfo {author} {\bibfnamefont {K.}~\bibnamefont
  {Funakubo}}\ and\ \bibinfo {author} {\bibfnamefont {E.}~\bibnamefont
  {Senaha}},\ }\href {\doibase 10.1103/PhysRevD.79.115024} {\bibfield
  {journal} {\bibinfo  {journal} {Phys. Rev. D}\ }\textbf {\bibinfo {volume}
  {79}},\ \bibinfo {pages} {115024} (\bibinfo {year} {2009})},\ \Eprint
  {http://arxiv.org/abs/0905.2022} {arXiv:0905.2022 [hep-ph]} \BibitemShut
  {NoStop}%
\bibitem [{\citenamefont {Weinberg}(1973)}]{Weinberg:1973am}%
  \BibitemOpen
  \bibfield  {author} {\bibinfo {author} {\bibfnamefont {E.~J.}\ \bibnamefont
  {Weinberg}},\ }\emph {\bibinfo {title} {{Radiative corrections as the origin
  of spontaneous symmetry breaking}}},\ \href@noop {} {Ph.D. thesis},\ \bibinfo
   {school} {Harvard U.} (\bibinfo {year} {1973}),\ \Eprint
  {http://arxiv.org/abs/hep-th/0507214} {arXiv:hep-th/0507214} \BibitemShut
  {NoStop}%
\bibitem [{\citenamefont {Jackiw}(1974)}]{Jackiw:1974cv}%
  \BibitemOpen
  \bibfield  {author} {\bibinfo {author} {\bibfnamefont {R.}~\bibnamefont
  {Jackiw}},\ }\href {\doibase 10.1103/PhysRevD.9.1686} {\bibfield  {journal}
  {\bibinfo  {journal} {Phys. Rev. D}\ }\textbf {\bibinfo {volume} {9}},\
  \bibinfo {pages} {1686} (\bibinfo {year} {1974})}\BibitemShut {NoStop}%
\bibitem [{\citenamefont {Dolan}\ and\ \citenamefont
  {Jackiw}(1974)}]{Dolan:1973qd}%
  \BibitemOpen
  \bibfield  {author} {\bibinfo {author} {\bibfnamefont {L.}~\bibnamefont
  {Dolan}}\ and\ \bibinfo {author} {\bibfnamefont {R.}~\bibnamefont {Jackiw}},\
  }\href {\doibase 10.1103/PhysRevD.9.3320} {\bibfield  {journal} {\bibinfo
  {journal} {Phys. Rev. D}\ }\textbf {\bibinfo {volume} {9}},\ \bibinfo {pages}
  {3320} (\bibinfo {year} {1974})}\BibitemShut {NoStop}%
\bibitem [{\citenamefont {Parwani}(1992)}]{Parwani:1991gq}%
  \BibitemOpen
  \bibfield  {author} {\bibinfo {author} {\bibfnamefont {R.~R.}\ \bibnamefont
  {Parwani}},\ }\href {\doibase 10.1103/PhysRevD.45.4695} {\bibfield  {journal}
  {\bibinfo  {journal} {Phys. Rev. D}\ }\textbf {\bibinfo {volume} {45}},\
  \bibinfo {pages} {4695} (\bibinfo {year} {1992})},\ \bibinfo {note}
  {[Erratum: Phys.Rev.D 48, 5965 (1993)]},\ \Eprint
  {http://arxiv.org/abs/hep-ph/9204216} {arXiv:hep-ph/9204216} \BibitemShut
  {NoStop}%
\bibitem [{\citenamefont {Nie}\ and\ \citenamefont {Sher}(1999)}]{Nie:1998yn}%
  \BibitemOpen
  \bibfield  {author} {\bibinfo {author} {\bibfnamefont {S.}~\bibnamefont
  {Nie}}\ and\ \bibinfo {author} {\bibfnamefont {M.}~\bibnamefont {Sher}},\
  }\href {\doibase 10.1016/S0370-2693(99)00019-2} {\bibfield  {journal}
  {\bibinfo  {journal} {Phys. Lett. B}\ }\textbf {\bibinfo {volume} {449}},\
  \bibinfo {pages} {89} (\bibinfo {year} {1999})},\ \Eprint
  {http://arxiv.org/abs/hep-ph/9811234} {arXiv:hep-ph/9811234} \BibitemShut
  {NoStop}%
\bibitem [{\citenamefont {Kanemura}\ \emph {et~al.}(1999)\citenamefont
  {Kanemura}, \citenamefont {Kasai},\ and\ \citenamefont
  {Okada}}]{Kanemura:1999xf}%
  \BibitemOpen
  \bibfield  {author} {\bibinfo {author} {\bibfnamefont {S.}~\bibnamefont
  {Kanemura}}, \bibinfo {author} {\bibfnamefont {T.}~\bibnamefont {Kasai}}, \
  and\ \bibinfo {author} {\bibfnamefont {Y.}~\bibnamefont {Okada}},\ }\href
  {\doibase 10.1016/S0370-2693(99)01351-9} {\bibfield  {journal} {\bibinfo
  {journal} {Phys. Lett. B}\ }\textbf {\bibinfo {volume} {471}},\ \bibinfo
  {pages} {182} (\bibinfo {year} {1999})},\ \Eprint
  {http://arxiv.org/abs/hep-ph/9903289} {arXiv:hep-ph/9903289} \BibitemShut
  {NoStop}%
\bibitem [{\citenamefont {Chen}\ \emph {et~al.}(2015)\citenamefont {Chen},
  \citenamefont {Dawson},\ and\ \citenamefont {Lewis}}]{Chen:2014ask}%
  \BibitemOpen
  \bibfield  {author} {\bibinfo {author} {\bibfnamefont {C.-Y.}\ \bibnamefont
  {Chen}}, \bibinfo {author} {\bibfnamefont {S.}~\bibnamefont {Dawson}}, \ and\
  \bibinfo {author} {\bibfnamefont {I.~M.}\ \bibnamefont {Lewis}},\ }\href
  {\doibase 10.1103/PhysRevD.91.035015} {\bibfield  {journal} {\bibinfo
  {journal} {Phys. Rev. D}\ }\textbf {\bibinfo {volume} {91}},\ \bibinfo
  {pages} {035015} (\bibinfo {year} {2015})},\ \Eprint
  {http://arxiv.org/abs/1410.5488} {arXiv:1410.5488 [hep-ph]} \BibitemShut
  {NoStop}%
\bibitem [{\citenamefont {Akeroyd}\ \emph {et~al.}(2000)\citenamefont
  {Akeroyd}, \citenamefont {Arhrib},\ and\ \citenamefont
  {Naimi}}]{Akeroyd:2000wc}%
  \BibitemOpen
  \bibfield  {author} {\bibinfo {author} {\bibfnamefont {A.~G.}\ \bibnamefont
  {Akeroyd}}, \bibinfo {author} {\bibfnamefont {A.}~\bibnamefont {Arhrib}}, \
  and\ \bibinfo {author} {\bibfnamefont {E.-M.}\ \bibnamefont {Naimi}},\ }\href
  {\doibase 10.1016/S0370-2693(00)00962-X} {\bibfield  {journal} {\bibinfo
  {journal} {Phys. Lett. B}\ }\textbf {\bibinfo {volume} {490}},\ \bibinfo
  {pages} {119} (\bibinfo {year} {2000})},\ \Eprint
  {http://arxiv.org/abs/hep-ph/0006035} {arXiv:hep-ph/0006035} \BibitemShut
  {NoStop}%
\bibitem [{\citenamefont {Aoki}\ \emph {et~al.}(2022)\citenamefont {Aoki},
  \citenamefont {Komatsu},\ and\ \citenamefont {Shibuya}}]{Aoki:2021oez}%
  \BibitemOpen
  \bibfield  {author} {\bibinfo {author} {\bibfnamefont {M.}~\bibnamefont
  {Aoki}}, \bibinfo {author} {\bibfnamefont {T.}~\bibnamefont {Komatsu}}, \
  and\ \bibinfo {author} {\bibfnamefont {H.}~\bibnamefont {Shibuya}},\ }\href
  {\doibase 10.1093/ptep/ptac068} {\bibfield  {journal} {\bibinfo  {journal}
  {PTEP}\ }\textbf {\bibinfo {volume} {2022}},\ \bibinfo {pages} {063B05}
  (\bibinfo {year} {2022})},\ \Eprint {http://arxiv.org/abs/2106.03439}
  {arXiv:2106.03439 [hep-ph]} \BibitemShut {NoStop}%
\bibitem [{\citenamefont {Chen}\ \emph {et~al.}(2020)\citenamefont {Chen},
  \citenamefont {Li},\ and\ \citenamefont {Wu}}]{Chen:2020wvu}%
  \BibitemOpen
  \bibfield  {author} {\bibinfo {author} {\bibfnamefont {N.}~\bibnamefont
  {Chen}}, \bibinfo {author} {\bibfnamefont {T.}~\bibnamefont {Li}}, \ and\
  \bibinfo {author} {\bibfnamefont {Y.}~\bibnamefont {Wu}},\ }\href {\doibase
  10.1007/JHEP08(2020)117} {\bibfield  {journal} {\bibinfo  {journal} {JHEP}\
  }\textbf {\bibinfo {volume} {08}},\ \bibinfo {pages} {117} (\bibinfo {year}
  {2020})},\ \Eprint {http://arxiv.org/abs/2004.10148} {arXiv:2004.10148
  [hep-ph]} \BibitemShut {NoStop}%
\bibitem [{\citenamefont {Haber}\ and\ \citenamefont
  {O'Neil}(2011)}]{Haber:2010bw}%
  \BibitemOpen
  \bibfield  {author} {\bibinfo {author} {\bibfnamefont {H.~E.}\ \bibnamefont
  {Haber}}\ and\ \bibinfo {author} {\bibfnamefont {D.}~\bibnamefont {O'Neil}},\
  }\href {\doibase 10.1103/PhysRevD.83.055017} {\bibfield  {journal} {\bibinfo
  {journal} {Phys. Rev. D}\ }\textbf {\bibinfo {volume} {83}},\ \bibinfo
  {pages} {055017} (\bibinfo {year} {2011})},\ \Eprint
  {http://arxiv.org/abs/1011.6188} {arXiv:1011.6188 [hep-ph]} \BibitemShut
  {NoStop}%
\bibitem [{\citenamefont {Haller}\ \emph {et~al.}(2018)\citenamefont {Haller},
  \citenamefont {Hoecker}, \citenamefont {Kogler}, \citenamefont {M\"onig},
  \citenamefont {Peiffer},\ and\ \citenamefont {Stelzer}}]{Haller:2018nnx}%
  \BibitemOpen
  \bibfield  {author} {\bibinfo {author} {\bibfnamefont {J.}~\bibnamefont
  {Haller}}, \bibinfo {author} {\bibfnamefont {A.}~\bibnamefont {Hoecker}},
  \bibinfo {author} {\bibfnamefont {R.}~\bibnamefont {Kogler}}, \bibinfo
  {author} {\bibfnamefont {K.}~\bibnamefont {M\"onig}}, \bibinfo {author}
  {\bibfnamefont {T.}~\bibnamefont {Peiffer}}, \ and\ \bibinfo {author}
  {\bibfnamefont {J.}~\bibnamefont {Stelzer}},\ }\href {\doibase
  10.1140/epjc/s10052-018-6131-3} {\bibfield  {journal} {\bibinfo  {journal}
  {Eur. Phys. J. C}\ }\textbf {\bibinfo {volume} {78}},\ \bibinfo {pages} {675}
  (\bibinfo {year} {2018})},\ \Eprint {http://arxiv.org/abs/1803.01853}
  {arXiv:1803.01853 [hep-ph]} \BibitemShut {NoStop}%
\bibitem [{\citenamefont {Wainwright}(2012)}]{Wainwright:2011kj}%
  \BibitemOpen
  \bibfield  {author} {\bibinfo {author} {\bibfnamefont {C.~L.}\ \bibnamefont
  {Wainwright}},\ }\href {\doibase 10.1016/j.cpc.2012.04.004} {\bibfield
  {journal} {\bibinfo  {journal} {Comput. Phys. Commun.}\ }\textbf {\bibinfo
  {volume} {183}},\ \bibinfo {pages} {2006} (\bibinfo {year} {2012})},\ \Eprint
  {http://arxiv.org/abs/1109.4189} {arXiv:1109.4189 [hep-ph]} \BibitemShut
  {NoStop}%
\bibitem [{\citenamefont {Belanger}\ \emph {et~al.}(2010)\citenamefont
  {Belanger}, \citenamefont {Boudjema}, \citenamefont {Pukhov},\ and\
  \citenamefont {Semenov}}]{Belanger:2010pz}%
  \BibitemOpen
  \bibfield  {author} {\bibinfo {author} {\bibfnamefont {G.}~\bibnamefont
  {Belanger}}, \bibinfo {author} {\bibfnamefont {F.}~\bibnamefont {Boudjema}},
  \bibinfo {author} {\bibfnamefont {A.}~\bibnamefont {Pukhov}}, \ and\ \bibinfo
  {author} {\bibfnamefont {A.}~\bibnamefont {Semenov}},\ }\href {\doibase
  10.1393/ncc/i2010-10591-3} {\bibfield  {journal} {\bibinfo  {journal} {Nuovo
  Cim. C}\ }\textbf {\bibinfo {volume} {033N2}},\ \bibinfo {pages} {111}
  (\bibinfo {year} {2010})},\ \Eprint {http://arxiv.org/abs/1005.4133}
  {arXiv:1005.4133 [hep-ph]} \BibitemShut {NoStop}%
\bibitem [{\citenamefont {Belanger}\ \emph {et~al.}(2021)\citenamefont
  {Belanger}, \citenamefont {Mjallal},\ and\ \citenamefont
  {Pukhov}}]{Belanger:2020gnr}%
  \BibitemOpen
  \bibfield  {author} {\bibinfo {author} {\bibfnamefont {G.}~\bibnamefont
  {Belanger}}, \bibinfo {author} {\bibfnamefont {A.}~\bibnamefont {Mjallal}}, \
  and\ \bibinfo {author} {\bibfnamefont {A.}~\bibnamefont {Pukhov}},\ }\href
  {\doibase 10.1140/epjc/s10052-021-09012-z} {\bibfield  {journal} {\bibinfo
  {journal} {Eur. Phys. J. C}\ }\textbf {\bibinfo {volume} {81}},\ \bibinfo
  {pages} {239} (\bibinfo {year} {2021})},\ \Eprint
  {http://arxiv.org/abs/2003.08621} {arXiv:2003.08621 [hep-ph]} \BibitemShut
  {NoStop}%
\bibitem [{\citenamefont {O'Hare}(2021)}]{OHare:2021utq}%
  \BibitemOpen
  \bibfield  {author} {\bibinfo {author} {\bibfnamefont {C.~A.~J.}\
  \bibnamefont {O'Hare}},\ }\href {\doibase 10.1103/PhysRevLett.127.251802}
  {\bibfield  {journal} {\bibinfo  {journal} {Phys. Rev. Lett.}\ }\textbf
  {\bibinfo {volume} {127}},\ \bibinfo {pages} {251802} (\bibinfo {year}
  {2021})},\ \Eprint {http://arxiv.org/abs/2109.03116} {arXiv:2109.03116
  [hep-ph]} \BibitemShut {NoStop}%
\bibitem [{\citenamefont {Kanemura}\ \emph {et~al.}(2005)\citenamefont
  {Kanemura}, \citenamefont {Okada},\ and\ \citenamefont
  {Senaha}}]{Kanemura:2004ch}%
  \BibitemOpen
  \bibfield  {author} {\bibinfo {author} {\bibfnamefont {S.}~\bibnamefont
  {Kanemura}}, \bibinfo {author} {\bibfnamefont {Y.}~\bibnamefont {Okada}}, \
  and\ \bibinfo {author} {\bibfnamefont {E.}~\bibnamefont {Senaha}},\ }\href
  {\doibase 10.1016/j.physletb.2004.12.004} {\bibfield  {journal} {\bibinfo
  {journal} {Phys. Lett. B}\ }\textbf {\bibinfo {volume} {606}},\ \bibinfo
  {pages} {361} (\bibinfo {year} {2005})},\ \Eprint
  {http://arxiv.org/abs/hep-ph/0411354} {arXiv:hep-ph/0411354} \BibitemShut
  {NoStop}%
\bibitem [{\citenamefont {Grojean}\ \emph {et~al.}(2005)\citenamefont
  {Grojean}, \citenamefont {Servant},\ and\ \citenamefont
  {Wells}}]{Grojean:2004xa}%
  \BibitemOpen
  \bibfield  {author} {\bibinfo {author} {\bibfnamefont {C.}~\bibnamefont
  {Grojean}}, \bibinfo {author} {\bibfnamefont {G.}~\bibnamefont {Servant}}, \
  and\ \bibinfo {author} {\bibfnamefont {J.~D.}\ \bibnamefont {Wells}},\ }\href
  {\doibase 10.1103/PhysRevD.71.036001} {\bibfield  {journal} {\bibinfo
  {journal} {Phys. Rev. D}\ }\textbf {\bibinfo {volume} {71}},\ \bibinfo
  {pages} {036001} (\bibinfo {year} {2005})},\ \Eprint
  {http://arxiv.org/abs/hep-ph/0407019} {arXiv:hep-ph/0407019} \BibitemShut
  {NoStop}%
\bibitem [{\citenamefont {Shiltsev}(2019)}]{Shiltsev:2019szl}%
  \BibitemOpen
  \bibfield  {author} {\bibinfo {author} {\bibfnamefont {V.}~\bibnamefont
  {Shiltsev}},\ }in\ \href@noop {} {\emph {\bibinfo {booktitle} {{17th
  Conference on Flavor Physics and CP Violation}}}}\ (\bibinfo {year} {2019})\
  \Eprint {http://arxiv.org/abs/1907.01545} {arXiv:1907.01545 [physics.acc-ph]}
  \BibitemShut {NoStop}%
\end{thebibliography}%

\end{document}